\documentclass[useAMS,usenatbib]{mn2e}
\usepackage{graphicx}
\usepackage{epsf}
\usepackage{bm}
\usepackage{color}
\usepackage{mathrsfs}
\usepackage{lscape}
\usepackage{times}
\usepackage{hyperref}
\usepackage{amsmath}
\usepackage{longtable}
\usepackage{tabularx}
\usepackage{booktabs}
\usepackage{threeparttable}
\usepackage{array}
\usepackage{lscape}
\usepackage{amssymb}
\newcommand\aap{{ A}\&{A}}

\newcommand\aj{{AJ}}
\newcommand\apj{{ApJ}}
\newcommand\apjl{{ApJ}}
\newcommand\apjs{{ApJS}}

\newcommand\mnras{{MNRAS}}

\newcommand\jcap{{JCAP}}

\newcommand\R{{(R)}}

\newcommand\vlos{$V_{\rm los}$}
\newcommand\vlosh{$V_{\rm los}^{\rm helio}$}
\newcommand\avlosh{$\overline{V}_{\rm los}^{\rm helio}$}

\title[Rotation curve \& mass distribution of the Milky Way] {The Milky Way's rotation curve out to 100 kpc and its constraint on the Galactic mass distribution}

\author[Y. Huang et al.]
               {Y. Huang$^{1}$\thanks{E-mails: yanghuang@pku.edu.cn (YH); x.liu@pku.edu.cn (XWL)}\thanks{LAMOST Fellow},
                X.-W. Liu$^{1,2}$\footnotemark[1],  H.-B. Yuan$^{3}$, M.-S. Xiang$^{4}$\footnotemark[2],  H.-W. Zhang$^{1,2}$, B.-Q. Chen$^{1}$\footnotemark[2]\\
                \newauthor J.-J. Ren$^{1}$, C. Wang$^{1}$, Y. Zhang$^{5}$, Y.-H. Hou$^{5}$, Y.-F. Wang$^{5}$, Z.-H. Cao$^{4}$ \\
$^{1}$Department of Astronomy, Peking University, Beijing, 100871, People's Republic of China\\
$^{2}$Kavli Institute for Astronomy and Astrophysics, Peking University, Beijing, 100871, People's Republic of China\\
$^{3}$Department of Astronomy, Beijing Normal University, Beijing, 100875, People's Republic of China\\
$^{4}$ Key Laboratory of Optical Astronomy, National Astronomical Observatories, Chinese Academy of Sciences, Beijing 100012, People's Republic of China\\
$^{5}$Nanjing Institute of Astronomical Optics \& Technology, National Astronomical Observatories, Chinese Academy of Sciences, Nanjing 210042}

\begin{document}

\date{}

\pagerange{\pageref{firstpage}--\pageref{lastpage}} \pubyear{2015}

\maketitle

\begin{abstract}
The rotation curve (RC) of the Milky Way out to $\sim$\,$100$\,kpc has been constructed using $\sim$\,$16,000$ primary red clump giants (PRCGs) in the outer disk selected from the LSS-GAC and the SDSS-III/APOGEE survey,  combined with $\sim$\,$5700$ halo K giants (HKGs) selected from the SDSS/SEGUE survey.  
To derive the RC, the PRCG sample of the warm disc population and the HKG sample of halo stellar population are respectively analyzed using a kinematical model allowing for the asymmetric drift corrections and re-analyzed using the spherical Jeans equation along with measurements of the anisotropic parameter $\beta$ currently available.
The typical uncertainties of RC derived from the PRCG and HKG samples are respectively $5$-$7$ km\,s$^{{-1}}$  and several tens km\,s$^{{-1}}$.
We determine a circular velocity at the solar position, {$V_c (R_0) = 240 \pm 6$\,km\,s$^{-1}$} and an azimuthal peculiar speed of the Sun, {$V_\odot = 12.1 \pm 7.6$\,km\,s$^{-1}$}, both in good agreement with the previous determinations.
The newly constructed RC has a generally flat value of $240$\,km\,s$^{-1}$ within a Galactocentric distance $r$ of 25 kpc and then decreases steadily to $150$\,km\,s$^{-1}$ at $r$\,$\sim$\,$100$\,kpc.
On top of this overall trend, the RC exhibits two prominent localized dips, one at $r$\,$\sim$\,$11$ kpc and another at $r$\,$\sim$\,$19$ kpc.
From the newly constructed RC, combined with other constraints, we have built a parametrized mass model for the Galaxy, yielding a virial mass of the Milky Way's dark matter halo of {$0.90^{+0.07}_{-0.08} \times 10^{12}$\,${\rm M}_{\odot}$} and a local dark matter density, { $\rho_{\rm \odot, dm} = 0.32^{+0.02}_{-0.02}$\,GeV\,cm$^{-3}$}.
\end{abstract}


\section{INTRODUCTION}
The rotation curve (hereafter RC) of the Milky Way gives the measured circular velocity $V_c$ as a function of the Galactocentric distance\,$r$.
The RC provides important constraints on the mass distribution of our Galaxy, including its dark matter (DM) content, as well as the local DM density (e.g. Salucci et al. 2010; Weber \& de Boer 2010). 
The latter is crucial for the interpretation of any signals that DM search experiments, direct or indirect, are expected to detect.
The RC can also be used to construct realistic Galactic mass model by fitting the RC with a parameterized multi-component Milky Way, consisting of, for instance, a bulge, a disc and a dark matter halo (e.g. Sofue, Honma, \& Omodaka 2009; Xin \& Zheng 2013).

Generally speaking, for the inner region (i.e. inside the solar circle) of the Galactic disc, the RC can be accurately measured simply using the so-called tangent-point (TP) method with the H~{\sc i}\,21\,cm or the CO\,2.6\,mm  gas emissions in the Galactic plane as tracer (Burton \& Gordon 1978; Gunn, Knapp \& Tremaine 1979; Clemens 1985; Fich, Blitz \& Stark 1989; Levine, Heiles, \& Blitz. 2008; Sofue et al.  2009).
In principle, a well defined RC could be established by this method for the entire Galactic inner region if one assumes that the gas moves in perfect circular orbits around the Galactic centre. 
However, the distribution and kinematics of gas can be easily perturbed by non-axisymmetric structures, in particular by the bar near the centre.
Given the presence of those perturbations, the TP method only works well in deriving the RC for the projected Galactocentric distance $R$ from $\sim$\,$4.5$\,kpc to $R_0$ (Galactocentric distance of the Sun; Chemin, Renaud \& Soubiran\,2015).
For the outer disc beyond the solar circle, the TP method can not be used to derive the RC. 
Instead, the RC is derived using a variety of tracers belonging to the cold disc populations from the measured line-of-sight velocities (\vlos) and estimated distances, such as the thickness of H~{\sc i} gas (Merrifield 1992; Honma \& Sofue 1997), H~{\sc ii} regions (Fich et al. 1989; Brand \& Blitz 1993; Turbide \& Moffat 1993), OB stars (Frink et al. 1996; Uemura et al. 2000; Bobylev \& Bajkova 2015), carbon stars (Demers \& Battinelli 2007; Battinelli et al. 2013) and classical cepheids (Pont et al. 1997).
However, two important issues limit the accuracy of RC derived from those disc tracers. 
First, it is difficult to determine the distances of those disc tracers and the poorly determined distances could lead large uncertainties (generally of the order of tens km\,s$^{-1}$, see, e.g. the Fig.\,1 of Sofue et al.  2009) in the derived circular velocity $V_c$.
Another issue is that, similar to the TP method, the underlying assumption that the disc tracers used move in purely circular orbits can be easily broken.
Disc tracers belonging to the cold populations, especially those young objects, are generally associated with the spiral arms and thus their kinematics are often perturbed by the arms.
At present, it is difficult to correct for the effects of those perturbations given the properties including dynamics of arms are still poorly understood.
Recently, accurate distances and values of \vlos\, have been measured for a number of masers\footnote{The masers are generally associated with young massive stars and compact H~{\sc ii} regions in the spiral arms.} by the Bar and Spiral Structure Legacy (BeSSeL) survey (Brunthaler et al. 2011), allowing, in principle, the determination of RC to a very high precision, say better than few km\,s$^{-1}$ (e.g. Xin \& Zheng 2013; Reid et al. 2014).
However, in deriving the RC from those measurements, possible perturbations to the measured velocities caused by the spiral arms remain to be properly accounted for.

For regions beyond the Galactic disc, the RC needs to be measured using halo stars, such as the blue horizontal branch (BHB) stars (Xue et al. 2008; Deason et al. 2012; Kafle et al. 2012; Williams \& Evans 2015) and the K giants (Bhattacharjee, Chaudhury, \& Kundu 2014, hereafter BCK14),  globular clusters or dwarf galaxies. 
For those tracers of halo populations (assuming isotropically distributed), the radial velocity dispersion $\sigma_r$, number density $\nu$ and velocity anisotropy parameter $\beta \equiv 1 - \sigma^2_t/\sigma^2_r$, are linked to the circular velocity $V_c$ through the Jeans equation (see, e.g. Binney \& Tremaine 2008, pp. 349) for spherical systems.
For halo tracers, profile of the radial velocity dispersion $\sigma_r$ can be easily estimated from the line-of-sight velocity dispersion $\sigma_{\rm los}$ (Battaglia et al. 2005; Dehnen et al. 2006), while their number density is found to follow a double power law with a break radius $r_{b}$ around $20$\,kpc (Bell et al. 2008; Watkins et al. 2009; Deason et al. 2011; Sesar et al. 2011).
However, the anisotropy parameter $\beta$ has only been accurately measured in the solar neighborhood, with a radially biased value between $0.5$\,--\,$0.7$ (Kepley et al. 2007; Smith et al. 2009; Bond et al. 2010;  Brown et al. 2010).
Due to the lack of accurate proper motion measurements of distant halo tracers, the anisotropy parameter $\beta$ is still poorly constrained beyond the solar neighborhood, particularly for the outer halo ($>$\,$25$\,kpc).
Hence, the existing determinations of RC suffer from the so-called {\it RC/mass--anisotropy degeneracy}.
To solve this problem, various values of the anisotropy parameter, either of arbitrary nature (e.g. BCK14) or predicted by numerical simulations (e.g. Xue et al. 2008; BCK14) have been adopted in the spherical Jeans equation to derive the RC. 
Only more recently, some constraints on the anisotropy parameter, mainly for the inner halo ($\leq$\,$25$\,kpc), have become available, based on some direct/indirect measurements (e.g. Deason et al. 2012; Kafle et al. 2012; Deason et al. 2013).

In this paper, we report a newly constructed RC of our Galaxy, the Milky Way, extending out to 100\,kpc, derived from $\sim$\,$16,000$ primary red clump giants (PRCGs) selected from the LAMOST Spectroscopic Survey of the Galactic Anti-centre (LSS-GAC; Liu et al. 2014; Yuan et al. 2015) and the SDSS-III/APOGEE survey (Eisenstein et al. 2011; Majewski et al. 2015) in the (outer) disc, as well as from $\sim$\,$5700$ halo K giants (HKGs) selected from the SDSS/SEGUE survey (Yanny et al. 2009) for the halo region. 
The usage of PRCGs in deriving the RC in the outer disc region help solve the above described two issues neatly. 
Firstly, PRCGs are considered as excellent standard candles given that their intrinsic luminosities are insensitive to the stellar populations (i.e. metallicity and age; e.g. Cannon 1970; Paczy{\'n}ski \& Stanek 1998).
Thus their distances can be determined to a much higher precision (typically 5--10\, percent) than for most other tracers belonging to cold disc population.
Secondly, PRCGs are of intermediate- to old-age stellar populations.
Thus they have enough time to dynamically mix in the disc and are therefore less affected by non-axisymmetric structures than those cold gaseous or young stellar tracers.
On the other hand, given that our PRCG sample stars are of relative old-age (i.e. warm), they need to be corrected the so-called asymmetric drifts (the offsets between the circular velocity and the mean rotational speed of the population concerned), which can be calculated from the velocity dispersions of our sample stars.
The large number of PRCGs employed in the current study dramatically reduces the random errors of the newly derived RC.
To derive the RC for the halo region, we have chosen SEGUE HKGs as tracers considering that, 1) They are intrinsically bright and also span about 4 mag in $r$-band absolute magnitude ($M_r$\,$\sim$\,$-1$ to $3$\,mag), allowing one to determine the RC out to a distance as far as 100\,kpc; 
2) They are abundantly observed in the SDSS/SEGUE survey.
We note that BCK14 have analyzed the same SEGUE HKG sample (and other two halo tracer samples) using the spherical Jeans equation and derived the RC in the halo region.
However, the analysis either assumes a constant anisotropy parameter $\beta$ or takes its value from numerical simulations, and thus could be liable to potential systematic uncertainties.
 To break the {\it RC/mass--anisotropy degeneracy}, we have re-analyzed the SEGUE HKG sample to derive the RC in the halo region using measurements of $\beta$ now available in the literature (see Section\,4.1).
Finally, we have constructed a new parameterized mass model for the Milky Way by combining constraints provided by the current, newly constructed RC and other available data.

The paper is organized as follows.
In Section\,2, we describe the LSS-GAC and SDSS data sets.
We derive the RC by modeling the PRCG and HKG samples in Sections\,3 and 4, respectively.
The combined, final RC out to 100\,kpc is presented in Section\,5.
A Galactic mass model derived by fitting the newly constructed RC is presented in Section\,6.
Finally, we summarize in Section\,7.

\section{Data}
\subsection{Coordinate systems}
In this study, we use three sets of coordinate systems: (1) A  right-handed  Cartesian system ($X, Y, Z$) positioned at  the Galactic centre with $X$ pointing in the direction opposite to the Sun, $Y$ in the direction of Galactic rotation and $Z$ towards the North Galactic Pole;  
(2) A  Galactocentric cylindrical system ($R, \phi, Z$) with $R$, the projected Galactocentric distance, increasing radially outward, $\phi$ in the direction of Galactic rotation and $Z$ the same as that in the Cartesian system;
(3) A  Galactocentric spherical coordinate system ($r, \theta, \phi$) with $r$, the Galactocentric distance, increasing radially outward, $\theta$ towards the Sourth Galactic Pole and $\phi$ in the direction of Galactic counter-rotation.
The Sun is assumed to be at the Galactic mid-plane (i.e. $Z = 0$\,pc) and has a value of $R_{0}$ of $8.34$\,kpc (Reid et al. 2014).
The former two coordinate systems are mainly used for disc stars and the spherical coordinate system is used for halo stars.
The three velocity components are represented by ($U, V, W$) in the Cartesian system centred on the Sun, ($V_{R}, V_{\phi}, V_{Z}$) in the Galactocentric cylindrical system and ($V_{r},  V_{\theta}, V_{\phi}$) in the Galactocentric spherical system.

\subsection{LSS-GAC, SDSS/SEGUE and SDSS-III/APOGEE data}
In this work, we use the second release of value-added catalogues of LSS-GAC (LSS-GAC DR2; Xiang et al. 2016, in preparation), the ninth SDSS/SEGUE public data release (SDSS/SEGUE DR9; Ahn et al. 2012) and the twelfth SDSS-III/APOGEE public data release (SDSS-III/APOGEE DR12; Alam et al. 2015).

LSS-GAC is a major component of the on-going LAMOST Experiment for Galactic Understanding and Exploration (LEGUE; Deng et al. 2012).
LSS-GAC aims to collect optical ($\lambda$$\lambda$3800--9000), low resolution ($R$\,$\sim$\,$1800$) spectra under dark and grey lunar conditions for a statistically complete sample of over three million stars of all colours and of magnitudes $14.0 \leq r < 17.8$ mag (18.5 mag for limited fields), in a continuous sky area of $\sim$\,$3400$ square degrees, centred on the GAC, covering Galactic longitudes $150 < l < 210^{\circ}$ and latitudes $|b| < 30^{\circ}$.
Over 2.5 million spectra of very bright stars ($9 < r < 14.0$ mag) in the equatorial Declination range $-10 < \delta < 60^{\circ}$ will also be obtained under bright lunar conditions.
The survey, initiated in the fall of 2012, is expected to last for five years.
Details about the survey, including the scientific motivations, target selections and data reduction, can be found in Liu et al. (2014) and Yuan et al. (2015).
The stellar atmospheric parameters and line-of-sight velocity  \vlos\, of LSS-GAC targets are derived with the LAMOST Stellar Parameter Pipeline at Peking University (LSP3; Xiang et al. 2015) using template matching with empirical spectral libraries.
LSP3 achieves an accuracy of 5.0 km\,s$^{-1}$, 150\,K, 0.25\,dex, 0.15\,dex for \vlos, effective temperature, surface gravity and metallicity [Fe/H], respectively, for spectra of FGK stars of signal-to-noise ratios (SNRs) per pixel at 4650\,\AA\,higher than 10.

SDSS/SEGUE survey, a Galactic extension of the SDSS-II/III surveys, has obtained a total of about 360,000 optical ($\lambda$$\lambda$3820--9100), low resolution ($R$\,$\sim$\,$2000$) spectra of Galactic  stars at different distances, from 0.5 to 100\,kpc (Yanny et al. 2009).
The spectra are processed with SEGUE Stellar Parameter Pipeline (SSPP;  Lee  et  al. 2008a,b;  Allende  Prieto  et  al. 2008; Smolinski et al. 2011), providing estimates of stellar parameters and \vlos.
The typical external errors of the stellar atmospheric  parameters yielded by SSPP are $\sim$\,$5$\,km\,s$^{-1}$ in \vlos, 180\,K in $T_{\rm eff}$, 0.24\,dex in $\log\,g$ and 0.23\,dex in [Fe/H] (Smolinski et al. 2011).
 
The SDSS-III/APOGEE survey collects high-resolution ($R$\,$\sim$\,$22,500$) and high SNRs ($\sim$\,$100$ per pixel) spectra in the near-infrared ($H$-band; 1.51 to 1.70 $\mu$m) for over one hundred thousand stars (mainly the red giant stars) in the Milky Way.
The scientific motivations and target selections are described in Majewiski et al. (2015) and Zasowski et al. (2013), respectively. 
The data reduction and stellar parameter determinations are introduced by Nidever et al. (2015) and Garc{\'i}a P{\'e}rez et al. (2015), respectively.
Calibrated with open clusters, the accuracy of APOGEE stellar parameters are better than 150\,K in $T_{\rm eff}$, 0.2\,dex in $\log\,g$ and 0.1\,dex in [Fe/H] (M{\'e}sz{\'a}ros et al. 2013). 
Benefited from the high resolution and high SNRs of APOGEE spectra, the random errors of \vlos\, delivered for APOGEE stars are at the level of $\sim$\,$0.1$\,km\,s$^{-1}$  with a zeropoint uncertainty at the level of $\sim$\,$0.5$\,km\,s$^{-1}$ (Nidever et al. 2015).

\subsection{PRCG and HKG samples}
\begin{figure}
\begin{center}
\includegraphics[scale=0.48,angle=0]{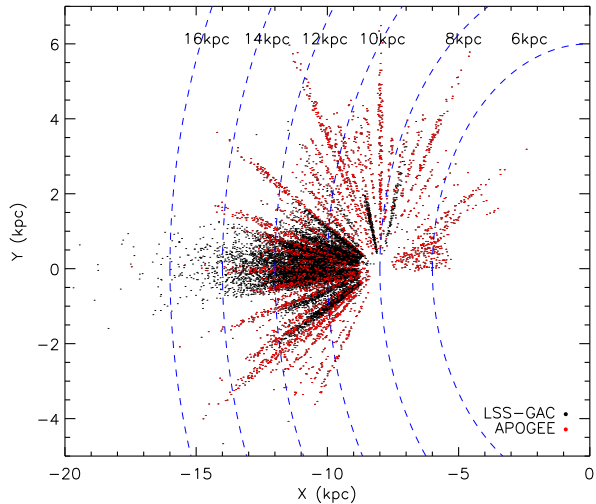}
\caption{Spatial distribution of PRCG sample stars in the $X$--$Y$ plane.
Black and red dots represent stars selected from LSS-GAC and APOGEE, respectively.
Blue dash lines denote different Galactocentric radii. }
\end{center}
\end{figure}

\begin{figure}
\begin{center}
\includegraphics[scale=0.42,angle=0]{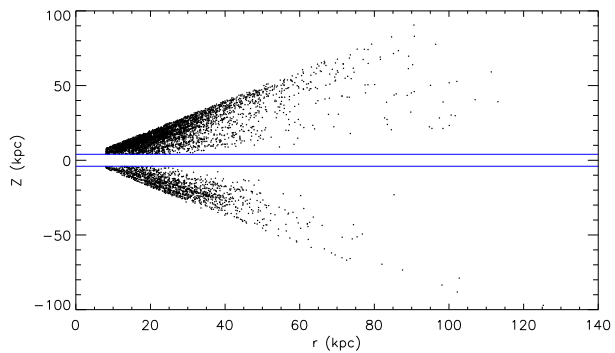}
\caption{Spatial distribution of  the SEGUE HKG sample stars in the $r$--$Z$ plane.
The two blue lines represent $|Z| = 4$\,kpc.}
\end{center}
\end{figure}

Specifically, as mentioned earlier, we use PRCGs selected from LSS-GAC and SDSS-III/APOGEE to derive the RC in the (outer) disc and HKGs selected from SDSS/SEGUE to derive that in the halo.
The PRCG stars are selected based on their positions in the metallicity dependent effective temperature-surface gravity and colour-metallicity stellar parameters spaces, as developed by Bovy et al. (2014) and applied to the APOGEE data.
From SDSS DR12, a total of 19,937 PRCGs are identified. 
Huang et al. (2015a) apply the same method  to the LSS-GAC DR2  and identify over 0.11 million PRCGs. 
The almost constant absolutes magnitude of PRCGs allow us to assign distances to the individual PRCGs with an accuracy of 5--10 per cent.  
For consistency of analysis, we have applied a zeropoint correction to \vlos\, values of LSS-GAC PRCG sample stars by adding a constant 2.7\,km\,s$^{-1}$  to those values.
This zeropoint offset between LSS-GAC and APOGEE $V_{\rm los}$ values is derived from a comparison of the two sets of measurements for  1500 common PRCG sample stars.
The correction is consistent with the finding of Xiang et al. (2015) who compare the values of \vlos\, for the LSS-GAC DR1 and APOGEE full samples that have about 3800 common sources.
To ignore the vertical motions in the following kinematic analysis and to minimize the contamination of halo stars,  we have restricted the PRCG sample to stars of  $|b| \leq 3^{\circ}$ and [Fe/H]\,$\geq -1.0$.
Finally, a total of 15,634 PRCGs are selected, with 11,572 stars from LSS-GAC and 3792 stars from APOGEE.
As Fig.\,1 shows, our PRCG sample spans from $R = 6$ to $16$\,kpc in the Galactic plane. 

The HKGs used here are taken from the SEGUE K giant catalog compiled by Xue et al. (2014).
The catalog  provides unbiased distance estimates with a typical precision of 16 per cent, as well as values of  \vlos\, and metallicities for a total of 6036 K giant stars.
To exclude possible contamination from the disc population, we have selected only those HKGs of $|Z| \geq 4$\,kpc from the catalog.
In addition, we cull of $r \leq 8.0$\,kpc  considering that only a few stars are found inside that radius.
Finally, a total of 5733 HKGs are selected.
As Fig.\,2 shows, the sample HKGs span a large range in Galactocentric radius $r$, from 8 to about 100\,kpc.

\section{RC from PRCGs}
\subsection{Kinematical model}
 \begin{figure}
\begin{center}
\includegraphics[scale=0.40,angle=0]{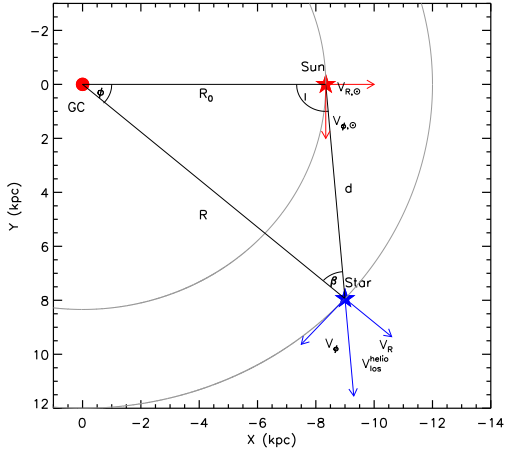}
\caption{Schematic diagram showing the observed heliocentric line-of-sight velocity of a star in the Galactic plane at position ($R$, $l$), where $R$ is the Galactocentric radius and $l$ the Galactic longitude.
The velocities are defined in cylindrical coordinates centred on the Galactic centre (GC).}
\end{center}
\end{figure}
For the PRCGs,  our approach to determine the RC is relied on the imprint that the Galactic rotation leaves in the observed heliocentric line-of-sight velocity \vlosh, as illustrated in Fig.\,3.
Specifically, the Galactic rotation yields a significant sinusoidal dependence on the Galactic longitude $l$ of the observed \vlosh\, at a certain value of $R$. 
Accordingly, one can derive the RC by fitting  the observed \vlosh\, as a function of $l$ of the PRCG  sample stars at different Galactocentric radii using a kinematical, axisymmetric model constructed as follows. 
Note that throughout the paper, the vertical motions are ignored and only those in the Galactic plane are considered since the sample includes only PRCGs of $|b|  \leq 3^{\circ}$ (see Section\,2.3).
As a result of the Galactic rotation, the average heliocentric line-of-sight velocities \avlosh\, of stars  at a given position ($R$, $l$) in the Galactic plane in the Galactocentric cylindrical frame is given by, 
\begin{equation}
             \begin{split}
              & \overline{{V}}_{\rm los}^{\rm helio}  = \overline{{V}}_{\phi} \R  \sin\beta - V_{\phi,\odot}\sin l\\ 
                 &\qquad \qquad + \overline{{V}}_{R} \R\cos\beta + {V}_{R,\odot}  \cos l ,
               \end{split}                                                                                                      
\end{equation}
where $\overline{{V}}_{\phi} \R = V_{c} \R - V_{a} \R$.
$V_{\phi,\odot}$ and ${V}_{R,\odot}$ are the Sun's azimuthal velocity and the radial component of its peculiar velocity, respectively.
$\overline{V}_{R}$ is the mean radial motion.
$\beta$ is the angle between the Sun and the Galactic centre with respect to the given position and the value of this angle is given by (see Fig.\,3),
\begin{equation}
\beta = \sin^{-1} (\frac{R_{0}}{R}\sin l).
\end{equation}
$V_{a} \R$ is the so-called asymmetric drift and is given by (e.g. Binney \& Tremaine 2008),
\begin{equation}
          \begin{split}
           & V_{a} \R  = \frac{\sigma_{R}^{2} \R }{2V_{c} \R} [\frac{\sigma_{\phi}^{2} \R}{\sigma_{R}^{2} \R}  - 1 + R (\frac{1}{R_{d}} + \frac{2}{R_{\sigma}}) \\
           &\qquad \qquad \qquad \qquad - \frac{R}{\sigma_{R}^{2} \R} \frac{\partial \overline{V_{R}V_{Z}}}{\partial Z}], 
          \end{split}
\end{equation}
assuming that both the number density $\nu$ of tracers and their (projected) radial velocity dispersion $\sigma_{R}$ are exponentially declining as a function of $R$ with scale lengths of $R_{\rm d}$ and $R_{\sigma}$, respectively.
The covariance $\overline{V_{R}V_{Z}}$ does not show obvious variations with $Z$ since our data are very close to the Galactic plane (B{\"u}denbender et al. 2015). 
Therefore, we can ignore the last term in the above Equation in the following analysis.

\begin{table}
\centering
\caption{Parameters of  the kinematical model employed }
\begin{threeparttable}
\begin{tabular}{lcc}
\hline
Parameter & Adopted value \\
\hline
$R_{\rm d}$ (kpc)& $2.5 \pm 0.5$\\

$\sigma_{\phi}^{2}$/$\sigma_{R}^{2}$ & $0.5 \pm 0.3$\\

$R_{\sigma}$ (kpc)& $16.40 \pm 1.25$\\

$\sigma_{R_{0}}$ (km\,s$^{-1}$)& $35.32 \pm 0.52$\\

$R_{0}$ (kpc)& $8.34 \pm 0.16$\\

$\Omega_{\odot}$ (km\,s$^{-1}\,{\rm kpc}^{-1}$)& $30.24 \pm 0.11$\\

$V_{\rm R,\odot}$ (km\,s$^{-1})$ &$ -7.01 \pm 0.20$\\
\hline
\end{tabular}
\begin{tablenotes}
\item[]
\end{tablenotes}
\end{threeparttable}
\end{table}

\begin{figure}
\begin{center}
\includegraphics[scale=0.5,angle=0]{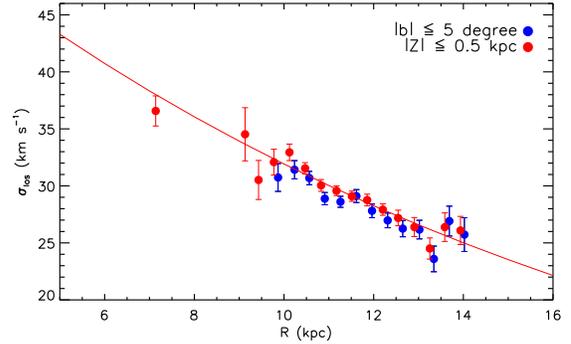}
\caption{Profile of line-of-sight velocity dispersion given by PRCG disc stars of [Fe/H]\,$\geq-1.0$ in the direction of Galactic centre ($l$\,$\sim$\,$0^{\circ}$) selected from APOGEE and in the direction of Galactic anti-centre ($l$\,$\sim$\,$180^{\circ}$) selected from LSS-GAC with the requirement of $|b| \leq 5^{\circ}$ (blue dots) or $|Z| \leq 0.5$\,kpc (red dots).
The red line represents an exponential best fit to the data points of requirement $|Z| \leq 0.5$\,kpc as described by Eq.\,(4).}
\end{center}
\end{figure}

\begin{figure*}
\begin{center}
\includegraphics[scale=0.6,angle=0]{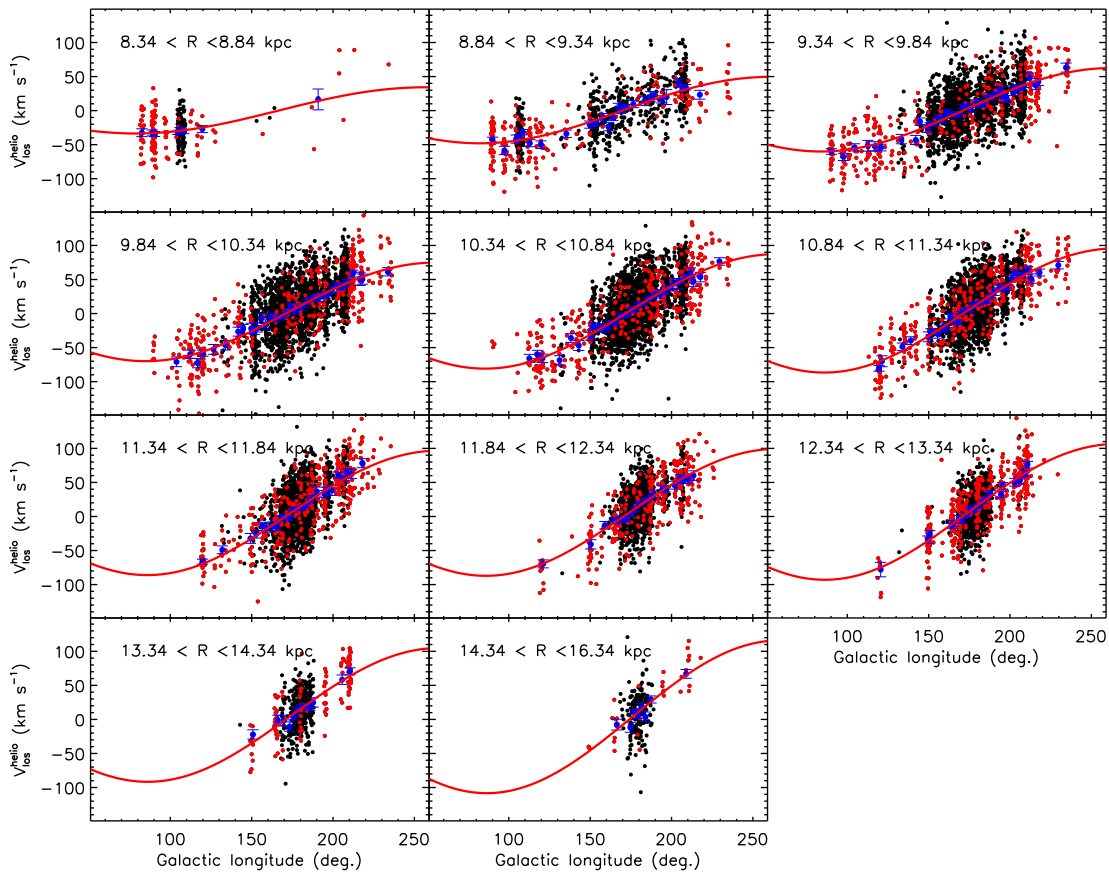}
\caption{Distributions of the heliocentric line-of-sight velocities (\vlosh) as a function of the Galactic longitude in the individual annuli of $R$ for the PRCG sample.
{Back and red dots represent LSS-GAC and APOGEE PRCGs, respectively.}
Blue dots in each annulus represent the mean heliocentric line-of-sight velocities of the individual Galactic longitude bins.
Red lines show the best fits to the data of the kinematical model described in the text.}
\end{center}
\end{figure*}

From Eq.\,(1), \avlosh\,of stars at a certain $R$ are essentially described by two trends of variations:
1) A sinusoidal dependence on the Galactic longitude governed by the mean Galactic rotation $\overline{V}_{\phi}$ of the (warm) stellar populations at the given $R$ and the azimuthal velocity of the Sun $V_{\phi,\odot}$;
2) A cosinusoidal dependence on the Galactic longitude governed by the mean radial motion\footnote{ Actually, \avlosh\, also varies with angle $\beta$ in a cosinusoidal form, as a result of the mean radial motion.
Given that $ \cos \beta = -\cos (\phi + l) \propto -\cos l$, there is also an approximately cosinusoidal  dependence of \avlosh on the Galactic longitude $l$ governed by the mean radial motion.} 
$\overline{V}_{R}$ of the (warm) stellar populations at the given $R$ and the Sun's peculiar velocity in the radial direction $V_{R,\odot}$.
For the first trend of variations, the mean Galactic rotation is a combination of the circular velocity (that we want to determine) and the asymmetric drift (that we need to correct for).
With $R_{d}$, $\sigma_{\phi}^{2} \R$/$\sigma_{R}^{2} \R$ and $\sigma_{R} \R$ known, the asymmetric drift $V_{a} \R$ becomes dependent on $V_{c}$ only [see Eq.\,(3)].
The scale length of the exponential disc, $R_{\rm d}$, has been studied extensively and is generally known as about 2.5\,kpc (e.g. Benjamin et al. 2005; Juri{\'c} et al. 2008).
For $\sigma_{\phi}^{2} \R$/$\sigma_{R}^{2} \R$, we assume it is independent of $R$ and has a fixed value of 0.5, approximately the mean value of existing measurements in the solar neighborhood (e.g. Dehnen \& Binney 1998a; Bovy et al. 2012).
The main unknown of Eq.\,(3) is $\sigma_R(R)$, i.e. the value of the exponentially declining radial velocity dispersion as a function of $R$.
Fortunately, from the existing data of APOGEE in the Galactic centre area (i.e. $l$\,$\sim$\,$0^{\circ}$) and those from LSS-GAC in the Galactic anti-centre area (i.e. $l$\,$\sim$\,$180^{\circ}$), we can measure the profile of $\sigma_{R}$ directly from the line-of-sight velocity dispersion $\sigma_{\rm los}$, since for those two areas, $\sigma_{R}$ is essentially identical to $\sigma_{\rm los}$.
For this  purpose, a total of $\sim$\,$4900$  PRCG disc stars of [Fe/H]\,$\geq -1.0$ are selected from LSS-GAC and APOGEE with $ |l-180| \leq 3.5^{\circ}$ for the area toward the Galactic anti-centre and with $|l| \leq 3.5^{\circ}$ for that toward the Galactic centre.
In doing so, we have also widened the cut on Galactic latitude by slightly, to $|b| \leq 5^{\circ}$ in order to include more stars.
Then we divide those stars into different bins in the radial direction and derive the line-of-sight velocity dispersion $\sigma_{\rm los}$ for each bin.
The binsize in the radial direction is allowed to vary to contain a sufficient number of stars in each bin.
 We require that the binsizes are no smaller than 0.3 kpc and each bin contains at least 80 stars.
As the blue dots in Fig.\,4 show,  $\sigma_{\rm los}$ shows a clear trend of declining with $R$.
The profile is not well constrained, given the limited range of  $R$  covered,  $10 \leq R \leq 14$\,kpc (no data points available from APOGEE).
To better constrain the profile, we have replaced the requirement  $|b| \leq 5^{\circ}$ with $|Z| \leq 0.5$\,kpc in selecting the stars.
For distant stars, the effect of the new requirement is similar to the original one, but it allows to include more nearby stars of relative high Galactic latitudes that are still close enough to the Galactic plane such that their vertical motions can be ignored.
Again, we derive $\sigma_{\rm los}$ by binning the stars in the radial direction and the results are overplotted in Fig.\,4 by red dots.
As expected, the new profile is similar to the original one for $R \geq 10$\,kpc, except that now it has data points in  the inner disc ($R$\,$\sim$\,$7$--10\,kpc).
To quantitively describe the profile of $\sigma_{R} $, we fit the measured data points of $\sigma_{los}$ obtained with the requirement of $|Z| \leq 0.5$\,kpc with an exponential function,
\begin{equation}
\sigma_{R} \R = \sigma_{R_{0}}exp(-\frac{R-R_{0}}{R_{\sigma}}),
\end{equation}
where $\sigma_{R_{0}}$ is the radial velocity dispersion at the solar position and $R_{\sigma}$ the scale length.
As shown by the red line in Fig.\,4, the best fit yields $\sigma_{R_{0}} = 35.32 \pm 0.52$\,km\,s$^{-1}$ and $R_{\sigma} = 16.40 \pm 1.25$\,kpc.
The value of $\sigma_{R_{0}}$ found here is consistent with the previous measurements for stars in the solar neighborhood (e.g. Dehnen \& Binney 1998a; Bensby et al. 2003; Soubiran et al. 2003).
The value of $R_{\sigma}$ agrees well with the recent determination of Sharma et al. (2014), who report $R_{\sigma}$\,$\sim$\,$14$\,kpc based on the RAVE (Steinmetz et al. 2006) data.

In principle, the mean radial motion $\overline{V}_{R}$ involved in the second trend of variations should be zero under our axisymmetric assumption.
However, based on the RAVE data, Siebert et al. (2011) and Williams et al. (2013) recently show that the mean radial motion $\overline{V}_{R}$ in the solar neighborhood is not zero and has a gradient in the radial direction.
To accomodate the possibility of a non-zero mean radial motion, we have left  $\overline{V}_{R}$ as a free parameter in our kinematical modeling.
We note that the effect of mean radial motion on \avlosh\, can be  easily disentangled from that of mean Galactic rotation considering that they have an opposite dependence on the Galactic longitude.
Finally, we fix the values of the azimuthal velocity of the Sun $V_{\phi,\odot}$ (involved in the first trend of variations), as well as the radial peculiar velocity,  $V_{R,\odot}$ (involved in the second trend of variations), using  the measurements in the literature. 
For $V_{\phi,\odot}$, it is identical to $\Omega_{\odot}R_{0}$.
As mentioned earlier, $R_{0}$ has been set to 8.34\,kpc (Reid et al. 2014).
The value of $\Omega_{\odot}$ is well constrained by the proper motions of Sgr\,$A^{*}$ measured  by Reid \& Brunthaler (2004).
For $V_{R,\odot}$, we take the value of $-7.01$\,km\,s$^{-1}$ determined by Huang et al. (2015b).
Table\,1 summaries all the fixed parameters employed in our kinematical model.

\begin{figure}
\begin{center}
\includegraphics[scale=0.42,angle=0]{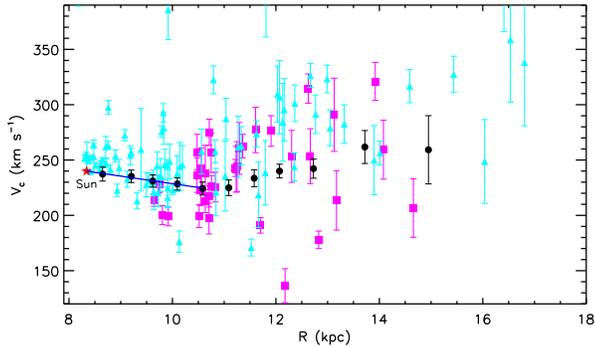}
\caption{Circular velocities of the Milky Way derived from our PRCG sample for the Galactocentric radius range $8\,\leq\,R\,\leq\,16$\,kpc (black dots).
Blue line is a linear fit to the RC for $R\,\leq\,11$\,kpc.
The red star denotes the circular velocity at the solar position as predicted by the linear fit.
Also overplotted cyan triangles and magenta boxes represent, respectively, measurements based H~{\sc ii} regions (Fich et al. 1989) and carbon stars of $60 \leq l \leq 150^{\circ}$ (Demers \& Battinelli 2007)}
\end{center}
\end{figure}

\begin{figure}
\begin{center}
\includegraphics[scale=0.42,angle=0]{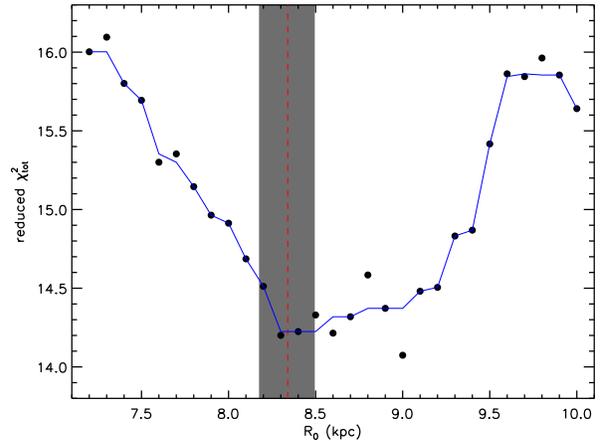}
\caption{$\chi^{2}_{\rm tot}$, sum of the reduced $\chi^{2}$ values of fit for the 11 annuli defined in Fig.\,5, as a function of the assumed value of $R_{0}$.
The blue line connecting the dots has been smoothed over three adjacent points.
Red dashed line represents the adopted value 8.34\,kpc of $R_{0}$ and the $1\,\sigma$ error of the adopted value of $R_{0}$ as estimated by Reid et al. (2014) is shown in grey shade.}
\end{center}
\end{figure}

\begin{figure}
\begin{center}
\includegraphics[scale=0.42,angle=0]{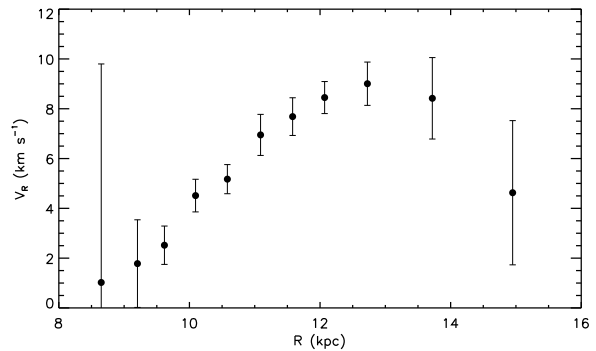}
\caption{Mean radial motion as a function of $R$ deduced from the PRCG sample.}
\end{center}
\end{figure}

\subsection{Fitting and results}
With the asymmetric drift properly modeled and a series of parameters fixed as described above, we are now left with two free parameters, i.e. the circular velocity $V_{c} \R$ and the mean radial motion $\overline{V}_{R} \R$,  to be determined by fitting \vlosh ($l$) measurements of our PRCG sample.
 To do so, we divide the stars into annuli in the radial direction with width 0.5\,kpc from $R_{0}$ to 12.34\,kpc, with width 1\,kpc from $12.34$ to $14.34$\,kpc and with width 2\,kpc for the last annulus, i.e. $14.34\,<\,R\,<\,16.34$\,kpc. 
 The choice of the width are consistent with the typical distance uncertainties of our PRCG sample (i.e. 5 per cent).
 For most annuli, the stars span from $\sim$\,$100$ to $210^{\circ}$ in Galactic longitude, wide enough to simultaneously obtain robust estimates of  $V_{c} \R$ and  $\overline{V}_{R} \R$.
Stars of $R\,\leq\,R_{0}$ in our sample are excluded given their narrow range of distribution in Galactic longitude.
 For each annulus, with the kinematical model described above, we fit the  \avlosh\, as a function of Galactic longitude.
To calculate the average heliocentric line-of-sight velocities at different Galactic longitudes, \avlosh\,$(l)$ measurements, we divide the stars of each annulus into bins of Galactic longitude. 
The binsize is allowed to vary but set to be no less than $2.5^{\circ}$ and each bin contains no less than 20 stars.
Finally, the best-fit values of $V_{c} \R$ and $\overline{V}_{R} \R$ of each annulus are found by nonlinear fitting that minimizes $\chi^{2}$ defined as,
\begin{equation}
\chi^{2} = \sum_{i=1}^{N} \frac{[\overline{V}_{\rm los, obs}^{\rm helio} (l_{i} , R) - \overline{V}_{\rm los, model}^{\rm helio} (l_{i},  R |\,\textbf{\emph p})]^{2}}{ \sigma_{\overline{V}_{\rm los, obs}^{\rm helio} (l_{i})}^{2}},
\end{equation}
where $N$ is the total number of data points to be fitted, $\sigma_{\overline{V}_{\rm los, obs}^{\rm helio} (l_{i})}$ is the uncertainty of  \avlosh\,$(l_{i})$, $l_{i}$ the mean longitude of the $ith$ Galactic longitude bin, and \textbf{\emph p} represents the parameters in the kinematical model [see Eq. (1)], including those of fixed values as listed in Table\,1 and, the circular velocity and mean radial motion to be derived from the fitting.

The fits are presented in Fig.\,5.
The derived RC, i.e., circular velocity $V_{\rm c}$ as a function of $R$, is presented in Fig.\,6.
{ To properly evaluate the errors of the derived $V_{c}$, we consider not only the fitting error $\sigma_{V_{c}}^{\rm fit}$ but also the error $\sigma_{V_{c}}^{\rm para}$ that propagates from the uncertainties of parameters fixed in the kinematical model (see Table 1) as measured by previous or current work.
Values of the latter are calculated using Monte Carlo simulations.
In practice, we obtain a distribution of the derived values of $V_{c}$ by repeating the fitting 5000 times, and infer the error $\sigma_{V_{c}}^{\rm para}$ from the distribution.
For each fit, values of those fixed parameters are randomly sampled assuming Gaussian distributions of values with uncertainties as listed in Table\,1.
The final error of the derived $V_{c}$ is then given by $\sqrt{(\sigma_{V_{c}}^{\rm fit})^{2} + (\sigma_{V_{c}}^{\rm para})^{2}}$.
The typical error of RC thus derived is only $5$-$7$ km\,s$^{-1}$. 
The relatively large errors (few tens km\,s$^{-1}$) in the few annuli of large $R$ are due to the relatively poor sampling in Galactic longitude for those annuli.
The newly derived RC shows a smooth trend of variations with $R$, except for a clear dip at $R$ around 11\,kpc.}
To infer the circular velocity at the solar position, $V_{c} (R_{0})$, we  apply a linear fit to the data point inside  the dip, i.e. $R \leqslant 11$\,kpc (Fig.\,6).
The fit yields {$V_{c} (R_{0}) = 239.89\,\pm\,5.92$\,km\,s$^{-1}$}, along with a local estimate of the slope of RC, i.e. {$\partial V_{c}$/$\partial R$, of $-6.85\,\pm\,3.90$\,km\,s$^{-1}$\,kpc$^{-1}$}.
Combining this estimate of $V_{c} (R_{0})$ and the known $V_{\phi, \odot}$, we find an azimuthal peculiar velocity of the Sun, {$V_{\odot}$, of $12.09\,\pm\,7.61$\,km\,s$^{-1}$}.
As defined in Section\,2.1, the current analysis assumes $R_{0} = 8.34$\,kpc as determined by Reid et al. (2014).
Actually, as evident from Eq.\,(1), the current data set also provide some constraints on $R_{0}$ and therefore can be used to check whether they are consistent with the adopted value of $R_{0}$.
For this purpose, we have calculated $\chi^{2}_{\rm tot}$, the sum of the reduced $\chi^{2}$ values of the fit for the 11 annuli defined in Fig.\,5, for various assumed values of $R_{0}$ ranging from 7.2 to 10.0\,kpc with a constant step 0.1\,kpc.
The results, plotted in Fig.\,7, show a clear minimum around 8.3\,kpc, identical to the value assumed above. 
Finally, we present the mean radial motions $\overline{V}_{R} \R$ deduced from the fitting in Fig.\,8.
The mean radial motion increases from $\sim$\,$1$\,km\,s$^{-1}$ at $R = 8.5$\,kpc to $\sim$\,$9$\,km\,s$^{-1}$ at $R = 12.5$\,kpc, and then decrease to $\sim$\,$4$\,km\,s$^{-1}$ at $R = 15$\,kpc. 
This trend of variations of $\overline{V}_{R}$ found here, together with a negative gradient of the radial motion found previously from the RAVE data for $7\,\lesssim\,R\,\lesssim\,8.5$\,kpc (Siebert et al. 2011; Williams et al. 2013), suggest that the value of mean radial motion oscillates with $R$.
This interesting result  is worth of further investigations but is out of the scope of the current study.

\subsection{Discussion}

\begin{figure}
\begin{center}
\includegraphics[scale=0.42,angle=0]{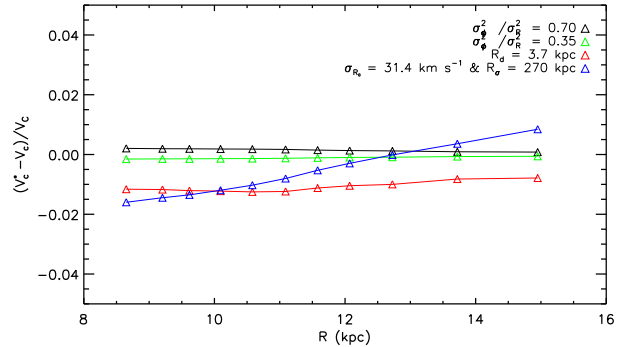}
\caption{Relative difference, $(V^{*}_{c} - V_{c})/V_{c}$, between the value of RC deduced by varying the assumed value of  parameters fixed in the kinematical model , i.e. $V^\ast_c$, and that derived assuming the canonical values of those parameters, i.e. $V_{c}$. 
Different colours of triangles represent the results for different sets of parameters as labelled in the top-right corner of the diagram.
Lines of different colours  are used to connect triangles of the same  colour to guide the eye. }
\end{center}
\end{figure} 

\subsubsection{Systematics}
To assess the systematic uncertainties of the newly derived RC, we first examine whether our choices of values of those fixed parameters in the kinematical model affect the results significantly.
To check the possible effects of $\sigma_{\phi}^{2}$/$\sigma_{R}^{2}$ on the derived RC, we have tried two values, i.e. a lowest value 0.35 and a highest value 0.70 reported in the literature, and redo the fitting.
The relative differences between the original RC, i.e. $V_c$, that derived assuming the canonical values of those parameters, and those derived after changing the canonical value of $\sigma_{\phi}^{2}$/$\sigma_{R}^{2}$ to the two extreme values above, denoted by $V^\ast_c$, are presented in Fig.\,9.
The differences are all smaller than 0.5 per cent ($\sim$\,$1.25$\,km\,s$^{-1}$).
{Similarly, by choosing a large scale length $R_{\rm d} = 3.7$\,kpc from Chang, Ko \& Peng (2011) and an almost flat radial velocity dispersion profile ($\sigma_{R_{0}} = 31.4$\,km\,s$^{-1}$ and $R_{\sigma}$ = 270\,kpc) from Bovy et al. (2012), we find that the resultant changes in our results (cf. Fig.\,9) are again very small, less than 2 per cent ($\sim$\,$5$\,km\,s$^{-1}$).}
As discussed in the above Section, our adopted value of $R_{0}$ is self-consistent with the current data set, suggesting our derived RC should suffer from negligible systematics, if any, as a result of our chosen value of  $R_{0}$.
Finally, at present, the best constraint on the $V_{\phi, \odot}$ comes from the proper motion measurements of Sgr\,$A^{*}$, yielding a value that is also in accordance with other most recent independent determinations (e.g. Bovy et al. 2012; Reid et al. 2014; Sharma et al. 2014).
Therefore, we have assumed that there are no systematic errors arising from our adopted value of $V_{\phi, \odot}$ in deriving the RC.
To conclude, the systematic errors of RC resultant as a consequence of our adopted canonical values of parameters are likely to be smaller than 2 per cent ($\sim$\,$5$\,km\,s$^{-1}$).

{ The kinematical model described above assumes a simplified Gaussian distribution of azimuthal velocity $V_{\phi}$.
In reality, for stars of warm population (such as those of our PRCG sample), the distribution of $V_{\phi}$ is significantly skewed (e.g. Sch{\"o}nrich \& Binney 2012, hereafter SB12).
However, the simplification is expected to have only minor effects on our derived RC.
As pointed out by Bovy et al. (2012), only at the tangent points the non-Gaussianity of $V_{\phi}$ distribution is fully visible as at those points the line-of-sight velocity is identical to $V_{\phi}$.
The PRCG sample employed in the current study are mostly spread over the Galactic longitudes between 90 and 240$^{\circ}$.
There is no tangent points in within this Galactic longitude range, hence the skewness of $V_{\phi}$ distribution is much less significant for oursample.  
To check the validity of the above simplification, we have further performed tests using mock-data, as presented in Appendix A. 
When generating the mock-data sets, the line-of-sight velocities are no longer sampled assuming a Gaussian distribution of $V_{\phi}$ but using a distribution calculated from the analytic formula given by SB12 (that well describes the skewness of $V_{\phi}$ distribution).
A detailed description of the mock-data sets and the  tests can be found in Appendix A.
We fit the mock-data sets using the above kinematical model (assuming a  simplified Gaussian distribution of $V_{\phi}$). 
The results show that the simplification does not introduce any significant bias in  the derived RC.   
}

In constructing the kinematical model, we have assumed an axisymmetric Galactic disc.
In reality, there are prominent non-axisymmetric structures in the Milky Way, such as the central bar and spiral arms, that may bias the derived RC.
 In principle, as discussed earlier, the PRCGs belong to relatively old and warm populations, and consequently should be relatively insensitive to perturbations by non-axisymmetric structures.
It is thus notable that, for the current PRCG sample, we have detected a non-zero mean radial motions, increasing from $\sim$\,$1$\,km\,s$^{-1}$ at $R = 8.5$\,kpc to $\sim$\,$9$\,km\,s$^{-1}$ at $R = 12.5$\,kpc, and then decreasing to $\sim$\,$4$\,km\,s$^{-1}$ at $R = 15$\,kpc (see Fig.\,8). 
Generally, non-axisymmetric structures could induce such streaming motions, both in the radial and azimuthal directions.
Therefore, the possibility that the derived RC is affected by the non-axisymmetric structures cannot be ruled out, considering that the trend seen in $\overline{V}_{R}$ may be caused by such perturbations.
By numerical simulations, one can quantitatively examine the effects of non-axisymmetric perturbations on warm populations, such as the PRCGs analyzed here, and figure out how large are the effects on the derived RC. 
However, we leave this to future work.

Finally, we note that the systematics caused by uncertainties in our  determinations of distance and \vlos\, of our PRCG sample stars are likely to be negligible, given  the high accuracies of the current distance and \vlos\,  estimates.

\subsubsection{Comparisons with other work}
As introduced, there have already many determinations of the RC in the disc based on data of various disc tracers of cold populations.
To compare those earlier results with ours, we have used the data (i.e. distances, radial velocities and errors) of H~{\sc ii} regions published by Fich et al. (1989) and those  of carbon stars of $60 \leq l \leq 150^{\circ}$ published by Demers \& Battinelli (2007), and  recalculate the circular velocities, adopting the same value of $R_{0}$ assumed in the current study, the value of $V_{c} (R_{0})$ deduced from the current study, as well as the solar peculiar velocities in the radial and vertical directions ($U_{\odot}$, $W_{\odot}$) from Huang et al. (2015b) for consistency reason.
The results are presented in Fig.\,6.
In generally, they show a trend of variations similar to our new measurements but with much large scatters.  
The large scatters are possibly due to the large distance errors in those data, or the perturbations of the non-axisymmetric structures, or both of them.
More recently, Bovy et al. (2012) derive the RC for $R$ from 4 to 14\,kpc with a kinematical model\footnote{We note that there is actually some differences between the kinematical analyses of  Bovy et al. (2012) and ours that their analysis has to assume a certain shape of RC (either flat or power-law) while ours does not.} quite similar to ours, using data of 3365 stars of warm populations selected from APOGEE.
In their analysis, they have assumed a flat and a power-law form of the RC when fitting the data.
Both approaches yield similar results and give $V_{c} (R_{0}) = 218 \pm 6$\,km\,s$^{-1}$.
This estimate of circular velocity at the solar position is substantially smaller than our value of {$V_{c} (R_{0}) = 239.89\,\pm\,5.92$\,km\,s$^{-1}$}, which is in excellent agreement with most of the recent independent determinations (e.g. McMillan 2011; Sch{\"o}nrich 2012; Reid et al. 2014; Sharma et al. 2014).
Their analysis also yields an estimate of $V_{\odot}$ of $26 \pm 3$\,km\,s$^{-1}$.
A very similar value, about 24\,km\,s$^{-1}$, is obtained by Bovy et al. (2015), using a sample of 8155 PRCG stars within 250\,pc from the Galactic mid-plane selected from the APOGEE, quite similar to the sample employed  in the current analysis.
Both estimates are more than 10\,km\,s$^{-1}$ higher than the values estimated from stars in the solar neighborhood (e.g. Sch{\"o}nrich, Binney \& Dehnen 2010; Huang et al. 2015b) as well as  the value deduced in the current study.
We note however, the analyses of both Bovy et al. (2012) and Bovy et al. (2015) assume a certain shape of the RC\,--\,a flat RC in fact. 
The analyses also adopt a flat radial velocity dispersion profile. 
Both assumptions have a major impact on  the estimated value of $V_{\odot}$ yet are not supported by the current data.

\subsubsection{The dip at $R$\,$\sim$\,$11$\,kpc}
Our newly derived RC shows a prominent dip at $R$\,$\sim$\,$11$\,kpc.
Such a similar feature (dip at $R$\,$\sim$\,$11$\,kpc) of RC has also been noted and studied by many previous studies (e.g. Sikivie 2003; Duffy \& Sikivie 2008; Sofue et al. 2009; de Boer \& Weber 2011).
Sikivie (2003) and Duffy \& Sikivie (2008) interpret the dip by the existence of hypothetical caustic rings of dark matter in the Galactic plane, with ring radii predicted at $a_{n} \simeq \frac{40 {\rm kpc}}{n}$ ($n = 1, 2, 3, 4,...$) for the Milky Way.
In this model, the $n =3$ hypothetical caustic ring is located at $a_{3} \simeq 13$\,kpc, and  is interpreted to be responsible for the dip of RC at $R\sim11$\,kpc.
Alternatively, de Boer \& Weber (2011) fit the dip with a donut-like ring of dark matter in the Galactic plane and conclude that the ring is actually at $R$\,$\sim$\,$12.4$\,kpc with a total mass of $\sim$\,$10^{10} {\rm M}_{\odot}$. 
We will address this feature as revealed by our more accurate RC  in a quantitatively way in Section\,6.
 Finally, we note that the possibility that some unknown perturbations, such as those induced by non-axisymmetric structures discussed above, are actually responsible for this localized dip at 13\,kpc, cannot be ruled out, considering that the derived RC is based on survey data that encompass only limited volumes of the whole Galaxy.

\section{RC from HKGs}
\subsection{Spherical Jeans model and results}
To derive the RC in the halo region, the spherical Jeans equation is applied to the HKG stars to estimate the circular velocity $V_{c} \R$ in equilibrium through the relation, 
\begin{equation}
V_{c}^{2} \R = - \sigma_{r}^{2} (\frac{d\,{\rm ln}\nu}{d\,{\rm ln}r} + \frac{d\,{\rm ln}\sigma_{r}^{2}}{d\,{\rm ln}r} + 2\beta),
\end{equation}
where $\sigma_{r}$ is the radial velocity dispersion of the HKGs and $\nu$ the number density of HKGs.
The velocity anisotropy parameter $\beta$ is defined as,
\begin{equation}
\beta = 1 - \frac{\sigma_{\theta}^{2} + \sigma_{\phi}^{2}}{2\sigma_{r}^{2}},
\end{equation}
where $\sigma_{\theta}$ and $\sigma_{\phi}$ are the polar and azimuthal velocity dispersions in the spherical coordinate system defined in Section\,2.1.

Recent extensive studies show that the number density of halo stellar population follows a broken power-law ($\nu \propto r^{-\alpha}$) distribution with a minor- to major-axis ratio $q = 0.5$--$1$, and a shallow slope of $\alpha$\,$\sim$\,$2$--$3$ out to a break radius $r_{b}$\,$\sim$\,$16$--27\,kpc followed by a steeper slope of $\alpha$\,$\sim$\,$3.8$--$5$ beyond $r_{b}$ (e.g. Bell et al. 2008; Watkins et al. 2009; Sesar et al. 2011, 2013; Deason 2011; Faccioli et al. 2014; Xue et al. 2015).
Similar to Kafle et al. (2014), here we adopt a spherical (i.e. $q = 1$), broken power-law distribution for the stellar halo, assuming  $\alpha = 2.4$ for the inner halo ($r \leq r_{b}$) and $\alpha = 4.5$ for the outer halo ($r > r_{b}$) from Watkins et al. (2009).
In agreement with the recent measurements (e.g. Bell et al. 2008; Sesar et al. 2013; Kafle et al. 2014; Xue et al. 2015), we set the broken radius $r_{b}$ to 20\,kpc.

As mentioned earlier, determinations of the RC for the halo region suffer from the so-called {\it RC--anisotropy degeneracy} because of the poorly constrained velocity anisotropy parameter $\beta$, especially for the outer halo.
There are however some recent progress in the measurements of $\beta$  to large distances, using direct or indirect methods that help break the degeneracy.
For the inner halo ($r \leq 12$\,kpc), $\beta$ is well, both directly and indirectly, measured and is found to have a radial biased value around 0.5 (e.g. Smith 2009; Brown 2010; Kalfe et a. 2012).
For the region $12 < r \leq 18$\,kpc, based on a sample of $\sim 4600$ BHB stars, Kafle et al. (2012) find that $\beta$ declines steadily, reaching a tangential value of $\sim -1.1$ at $\sim 17$\,kpc.
From the proper motions of main-sequence stars measured by {\it HST}, Deason et al. (2013) find the halo is isotropic ($\beta = 0.0^{+0.2}_{-0.4}$) at $r = 24 \pm 6$\,kpc.
The halo beyond  $50$\,kpc is again found to be radially biased with $\beta = 0.4 \pm 0.2$, based on the most recent study of Kafle et al. (2014). 
Table\,2 summarizes these latest measurements of $\beta$.
The Table is used to infer $\beta$ at any given $r$ by interpolation in our following analysis. 

\begin{table}
\centering
\caption{The velocity anisotropy parameter $\beta$ at different radii}
\begin{threeparttable}
\begin{tabular}{cc}
\hline
$r$&$\beta$\\
(kpc)&\\
\hline
$[8.0,12.0]$&$+0.50^{+0.10}_{-0.10}$\\
13.2&$+0.47^{+0.23}_{-0.29}$\\
14.0&$+0.21^{+0.32}_{-0.23}$\\
15.1&$-0.23^{+0.51}_{-0.60}$\\
16.1&$-0.64^{+0.68}_{-0.87}$\\
16.9&$-1.08^{+0.78}_{-1.01}$\\
17.9&$-0.62^{+0.74}_{-0.98}$\\
$[18.0,30.0]$&$+0.00^{+0.20}_{-0.40}$\\
$> 50$&$+0.40^{+0.20}_{-0.20}$\\
\hline
\end{tabular}
\begin{tablenotes}
\item[]
\end{tablenotes}
\end{threeparttable}
\end{table}

Finally, we determine the last unknown term in the Jeans equation -- the profile of radial velocity dispersion. 
To do so, we first convert the observed \vlosh to the Galactic standard of rest (GSR) frame by,
\begin{equation}
V_{\rm GSR} = V_{\rm los}^{\rm helio}  + U_{\odot}\cos b \cos l + V_{\phi, \odot}\cos b \sin l + W_{\odot}\sin b,
\end{equation}
where $U_{\odot}$ and $W_{\odot}$ are taken from Huang et al. (2015b), and $V_{\phi, \odot}$ is again set to the value as adopted in Section\,3.1.
Next, we calculate the GSR line-of-sight velocity dispersion, $\sigma_{\rm GSR}$, by dividing HKGs into different radial bins.
The binsize in radial direction is allowed to vary  such that each bin contains at least 40 stars.
We require that the radial binsizes are no smaller than 1.0\,kpc for $r \leq 20$\,kpc, 2.5\,kpc for $20 < r \leq 50$\,kpc and 5.0\,kpc for $r > 50$\,kpc, respectively, to match with the typical distance uncertainties of our HKG sample stars (i.e. 16 per cent).
The results are presented in Fig.\,10.
Then values of the radial velocity dispersion $\sigma_{r}$ can be obtained by applying a correction factor (Dehnen et al. 2006) to $\sigma_{\rm GSR}$,
\begin{equation}
\sigma_{r} = \frac{\sigma_{\rm GSR}}{\sqrt{1 - \beta A(r)}},
\end{equation}
where
\begin{equation}
A(r) = \frac{r^{2} + R_{0}^{2}}{4r^{2}} - \frac{(r^{2} - R_{0}^{2})^{2}}{8r^{3}R_{0}}{\rm ln}|\frac{r+R_{0}}{r-R_{0}}|.
\end{equation}
The profile of $\sigma_{r}$, estimated from the above equations, is also presented in Fig.\,10.
{ The uncertainty of $\sigma_{\rm GSR}$ for each bin is estimated by the classical method, $\Delta\sigma_{\rm GSR} = \sqrt{1/[2(N-1)]}\sigma_{\rm GSR}$, where $N$ is the total number of stars in the bin.
Then the uncertainty of $\sigma_{r}$ for each bin is propagated from that of $\sigma_{\rm GSR}$ using Eq.\,(9).
To account for possible uncertainties induced by the correction factor for $\sigma_{r}$, we run Monte Carlo simulations.
This is realized using the uncertainties of the input quantities ($\beta$, $r$ and $\sigma_{\rm GSR}$) in Eq.\,(9) for a given bin and randomly sampling those  quantities assuming Gaussian error distributions.
For each bin,  we obtain a distribution of $\sigma_{r}$ by repeating the sampling 1000 times, and infer the error of $\sigma_{r}$ for that bin from the distribution.}
 Similar to the density profile, the profile of $\sigma_{r}$ also shows a broken radius around 20\,kpc, with a steeper slope inside the broken radius than beyond.
To better describe the profile, we apply a double power-law fit ($\sigma_{r} \propto r^{-\gamma}$) to the profile with a broken radius of 20\,kpc.
The fit yields an inner slope of $\gamma = 0.43$ and an outer slope of $\gamma = 0.24$. 

\begin{figure}
\begin{center}
\includegraphics[scale=0.42,angle=0]{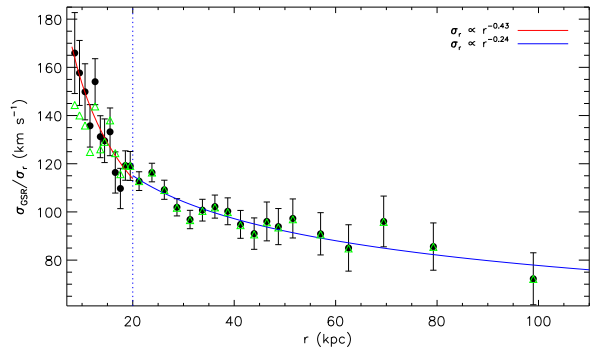}
\caption{Velocity dispersions derived from the HKG sample. 
Green triangles  and black circles represent the values of line-of-sight velocity dispersion in the GSR frame $\sigma_{\rm GSR}$, and those of radial velocity dispersions $\sigma_r$, respectively.
The dotted line indicates the broken radius ($r = 20$\,kpc) of the $\sigma_r$ profile.
Red and blue lines show the best power-laws fits to the profile within and beyond the broken radius, respectively.}
\end{center}
\end{figure} 

\begin{figure}
\begin{center}
\includegraphics[scale=0.42,angle=0]{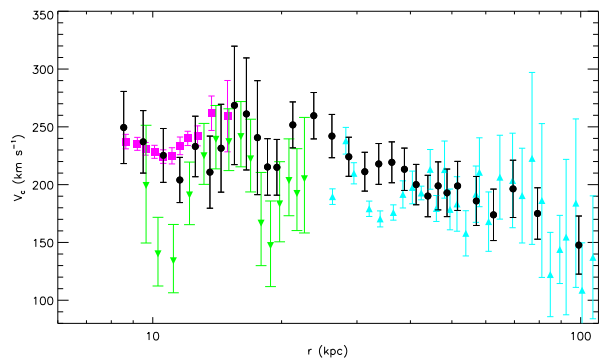}
\caption{Circular velocities of the Milky Way derived from the HKG sample for the range $8 \leq r \leq 100$\,kpc (black circles).
Magenta boxes are values derived from the PRCG sample.
Also overplotted as green downward triangles and cyan triangles are, respectively, determinations taken from Kafle et al. (2012) and BCK14.}
\end{center}
\end{figure} 

We now can obtain the RC by solving the spherical Jeans Eq.\,(6) with values of $\nu$, $\beta$ and $\sigma_{r}$ for the HKG sample properly specified above.
The derived RC is presented in Fig.\,11.
The uncertainty of circular velocity $V_{c}$ in each bin is again calculated using a Monte Carlo approach similar to that described above.
The typical errors of the RC are several tens km\,s$^{-1}$.
The largest errors found at radii around $20$\,kpc are due to the poorly constrained velocity anisotropy parameter $\beta$ around that region.
In the region $8 \leq r \leq 25$\,kpc, the RC shows two localized dips at radii $\sim$\,$11$ and $\sim$\,$19$\,kpc, respectively.
The inner one is exactly that already found above from the PRCG sample.

\subsection{Discussion}
\subsubsection{Systematics}
As evident from Eq.\,(6), varying the power-law indexes of stellar density of the inner and/or the outer halo change the circular velocities derived systematically.
To explore the sensitivity of the newly derived RC to the density profile adopted, we have repeated the analysis using another two sets of indexes, corresponding, respectively, the lower and upper limits allowed by the current available measurements. 
For the lower limit, we use a power-law index of 2.0 and 3.8 for the inner and outer halo, respectively.
We find that the circular velocities thus derived are on average 8 per cent smaller than the original values in both the inner and outer halo regions.
 For the upper limit, we adopt a power-law index of 3.0 and 5.0 for the inner and outer halo, respectively, and find that the circular velocities deduced in the inner and outer halo regions are, respectively, 9 and 6 per cent larger than the original ones on average. 
Thus the errors of the newly derived RC resultant from  the possible uncertainties in density profile  are probably less than 10 per cent as a whole (i.e. $\sim$\,$20$\,km\,s$^{-1}$), which are comparable to the random errors of the newly derived RC.

\begin{table*}
\centering
\caption{Final combined RC of the Milky Way}
\begin{threeparttable}
\begin{tabular}{lccc|lccc}
\hline
\hline
$r$&$V_{c}$&$\sigma_{V_{c}}$&tracer&$r$&$V_{c}$&$\sigma_{V_{c}}$&tracer\\
(kpc)&(km\,s$^{-1}$)&(km\,s$^{-1}$)&&(kpc)&(km\,s$^{-1}$)&(km\,s$^{-1}$)&\\
\hline
 4.60&231.24& 7.00&  H~{\sc i}&17.56&240.66&49.91&  HKG\\
 5.08&230.46& 7.00&  H~{\sc i}&18.54&215.31&24.80&  HKG\\
 5.58&230.01& 7.00&  H~{\sc i}&19.50&214.99&24.42&  HKG\\
 6.10&239.61& 7.00&  H~{\sc i}&21.25&251.68&19.50&  HKG\\
 6.57&246.27& 7.00&  H~{\sc i}&23.78&259.65&19.62&  HKG\\
 7.07&243.49& 7.00&  H~{\sc i}&26.22&242.02&18.66&  HKG\\
 7.58&242.71& 7.00&  H~{\sc i}&28.71&224.11&16.97&  HKG\\
 8.04&243.23& 7.00&  H~{\sc i}&31.29&211.20&16.43&  HKG\\
 8.34&239.89& 5.92&  MRCG&33.73&217.93&17.66&  HKG\\
 8.65&237.26& 6.29&  MRCG&36.19&219.33&18.44&  HKG\\
 9.20&235.30& 5.60&  MRCG&38.73&213.31&17.29&  HKG\\
 9.62&230.99& 5.49&  MRCG&41.25&200.05&17.72&  HKG\\
10.09&228.41& 5.62&  MRCG&43.93&190.15&18.65&  HKG\\
10.58&224.26& 5.87&  MRCG&46.43&198.95&20.70&  HKG\\
11.09&224.94& 7.02&  MRCG&48.71&192.91&19.24&  HKG\\
11.58&233.57& 7.65&  MRCG&51.56&198.90&21.74&  HKG\\
12.07&240.02& 6.17&  MRCG&57.03&185.88&21.56&  HKG\\
12.73&242.21& 8.64&  MRCG&62.55&173.89&22.87&  HKG\\
13.72&261.78&14.89&  MRCG&69.47&196.36&25.89&  HKG\\
14.95&259.26&30.84&  MRCG&79.27&175.05&22.71&  HKG\\
15.52&268.57&49.67&  HKG&98.97&147.72&23.55&  HKG\\
16.55&261.17&50.91&  HKG&--&--&--&--\\
\hline
\end{tabular}
\begin{tablenotes}
\item[]
\end{tablenotes}
\end{threeparttable}
\end{table*}

In the current study, we assume that all the HKGs in our sample are of a single halo population in the Jeans equation.
Recently, some studies (Kafle et al. 2013; Hattori et al. 2013) claim that there is a correlation between the metallicity and kinematics of halo stars such that metal-rich ([Fe/H]$ > -2$) and metal-poor ([Fe/H]$ < -2$) halo stars may actually belong to different populations.
The correlation, if exists, may potentially affect our analysis.
On the other hand, a revisit of the problem by Fermani and Sch{\"o}rich (2013) find no correlation at all.
In the future, with even larger halo samples than the current available, it is possible to examine this effect quantitatively by modeling metal-rich and metal-poor populations separately using the Jeans equation. 

Finally, we note that the current analysis assumes a spherical stellar halo.
It may well be that the stellar halo is not spherical.
We will however leave the determination of the RC in a non-spherical halo to future studies.

\subsubsection{Comparisons with other work}
We first compare the circular velocities  derived from the HKG sample to those from the PRCG sample in the overlap region $8 \leq r \leq 16$\,kpc. 
As Fig.\,11 shows, they are in good agreement within the errors and the dip at $r$\,$\sim$\,$11$\,kpc discussed above  is again revealed by data from the HKG sample.  
The close agreement between the two sets of independent determinations suggests the robustness of our analysis based on two types of tracer of different populations.
Recently, Kafle et al. (2012) estimate the RC in the region $8 \leq r \leq 25$\,kpc with the spherical Jeans equation using 4664 BHB stars.
Their results are overplotted in Fig.\,11.
Again, as Fig.\,11 shows, the measurements also show two prominent dips at $r$\,$\sim$\,$11$ and $\sim$\,$19$\,kpc,  consistent with our result.
We notice that the two dips revealed by their data are much deeper than ours, especially for the first dip. 
The discrepancies are largely caused by the differences in the radial velocity profiles between their BHB and our HKG samples.
As mentioned earlier, BCK14 have recently derived the RC in the outer halo of $25 \leq r \leq 200$\,kpc, also with the spherical Jeans equation, using three halo tracer samples, including the HKG sample used in the current work.
We overplot in Fig.\,11 their values that are the combined results from the three halo tracer samples deduced by setting $R_{0} = 8.3$\,kpc, $V_{c} (R_{0}) = 244$\,km\,s$^{-1}$, and assuming a $\beta$ profile taken from the numerical simulation of Rashkov et al. (2013).
As one can see in Fig.\,11, their circular velocities are somewhat smaller than ours for $r \leq 30$\,kpc. 
Beyond this, they are in good agreement.
The discrepancies inside $30$\,kpc are largely due to the differences of $\beta$ adopted -- they use $\beta$\,$\sim$\,$0.6$ as given by the simulation while we set it to 0.0 based on the direct measurement of Deason et al. (2013).
For $r \ge 30$\,kpc,  the values of $\beta$ given by the simulation, between 0.50\,--\,0.75, are close to the value of 0.4 adopted by us.
Thus it is no surprising that the sets of two RC agree in general within the errors.

\subsubsection{The dip at $r$\,$\sim$\,$19$\,kpc}
In addition to the significant, localized dip at $r$\,$\sim$\,$11$\,kpc discussed in Section\,3.3.3, another prominent dip at $r$\,$\sim$\,$19$\,kpc is revealed in the RC derived by the current HKG sample.
As mentioned above, the latter is also seen in the RC derived by Kafle et al. (2012) using over 4000 BHB stars.
{ Recall that the profiles of stellar number density, velocity anisotropy and radial velocity dispersion,  described in Section\,4.1, all have a break around $20$\,kpc.
Thus it is tempting to conjecture that the dip seen in RC at $r$\,$\sim$\,$19$\,kpc is a direct consequence of the breaks in those profiles that may all have a common cause.
On the other hand, the possibility that the dip is artificial cannot be completely ruled out,  given the current measurement uncertainties of the break radii as well as the slopes of those profiles.
Finally, note that the position of the dip at $r$\,$\sim$\,$19$\,kpc coincides roughly with an $n = 2$ ring radius, $a_2 \simeq 20$\,kpc, of a hypothetical caustic ring of dark matter  in the Galactic plane proposed by Sikivie (2003) and Duffy \& Sikivie (2008). 
In Section\,6, we will present further quantitive analysis of this dip.

\section{Final combined RC}
In this Section, we combine the two segments of RC derived above from, respectively, a sample of PRCGs selected from LSS-GAC and  APOGEE (Fig.\,6) and from a sample of HKGs selected from  SEGUE (Fig.\,11).  
For the overlap region (i.e. $8 \leq r \leq 15$\,kpc) of the two segments, the circular velocities derived from the PRCGs are adopted as the final values given their high accuracy almost an order of magnitude higher than those derived from the HKGs.
In addition, to provide circular velocities for the inner disk region (inside the solar circle), we take the H~{\sc i} measurements of Fich et al. (1989) based on the TP method  described above.
We only provide  H~{\sc i} data for the region between $\sim$\,$4.5$\,kpc  and $R_{0}$, believed to be less affected by the non-axisymmetric structures (e.g. the central bar; Chemin et al. 2015).  
For consistency, we have recalculated the circular velocities from those H~{\sc i} data adopting $R_{0} = 8.34$\,kpc and  $V_{c} (R_{0}) = 239.89$\,km\,s$^{-1}$ described above.
The circular velocities for the inner disk region are provided by taking the mean of circular velocities derived from the H~{\sc i} data in every 0.5 kpc radial bin.
The uncertainties of the mean circular velocities are assumed to be { 7.0}\,km\,s$^{-1}$. 
The final combined values of circular velocity at different radius $r$, their associated $1\sigma$ errors ($\sigma_{V_{c}}$) and the tracer used, are presented in Table\,3. 
This final combined RC is plotted in Fig.\,12.
Generally, the combined RC has a flat value $ 240$\,km\,s$^{-1}$ within $r$\,$\sim$\,$25$\,kpc.
Beyond this, it starts to decline steadily, reaching $150$\,km\,s$^{-1}$ at $r$\,$\sim$\,$100$\,kpc.  
In addition to the overall trend, two prominent localized dips, as described earlier, are clearly seen in the RC, with one at $r$\,$\sim$\,$11$\,kpc and another at $r$\,$\sim$\,$19$\,kpc.

\section{Galactic mass models based on the combined RC}
\subsection{Galactic mass models and the fit results}
Modeling the mass distribution of the Milky Way is a fundamental task of Galactic astronomy (e.g. Dehnen \& Binney 1998b, hereafter DB98b; Klypin, Zhao \& Somerville 2002).
It is also of vital importance for understanding the Galaxy formation and evolution, bearing fundamental questions such as whether the Galactic disc is maximal (e.g. Sackett 1997), whether there is a `missing baryon problem' in our Galaxy (e.g. Klypin et al. 1999) and whether our Galaxy is an archetypical spiral galaxy comparing to other local spiral galaxies (e.g. Hammer et al. 2007).
 As introduced in Section\,1, the RC provides the most fundamental, direct probe of the  mass distribution of the Milky Way.  
In doing this, we have constructed a parametrized Galactic mass model by fitting the model predicted RC to our newly derived combined one presented above.
This parametrized Galactic mass model consists of four major components, i.e. three discs, a bulge, a dark matter halo and two rings. 
The model predicted circular velocities as a function of Galactic radius are contributed by the four components as given by,
\begin{equation}
V_{c}^{2} = {V_{c, \rm disc}^{2} + V_{c, \rm bulge}^{2} + V_{c, \rm halo}^{2} +  V_{c, \rm ring}^{2}}. 
\end{equation}
Except for the rings, we adopt density profiles of the other three major components similar to those employed by DB98b and McMillan (2011).
They are briefly describe below.

\begin{figure*}
\begin{center}
\includegraphics[scale=0.82,angle=0]{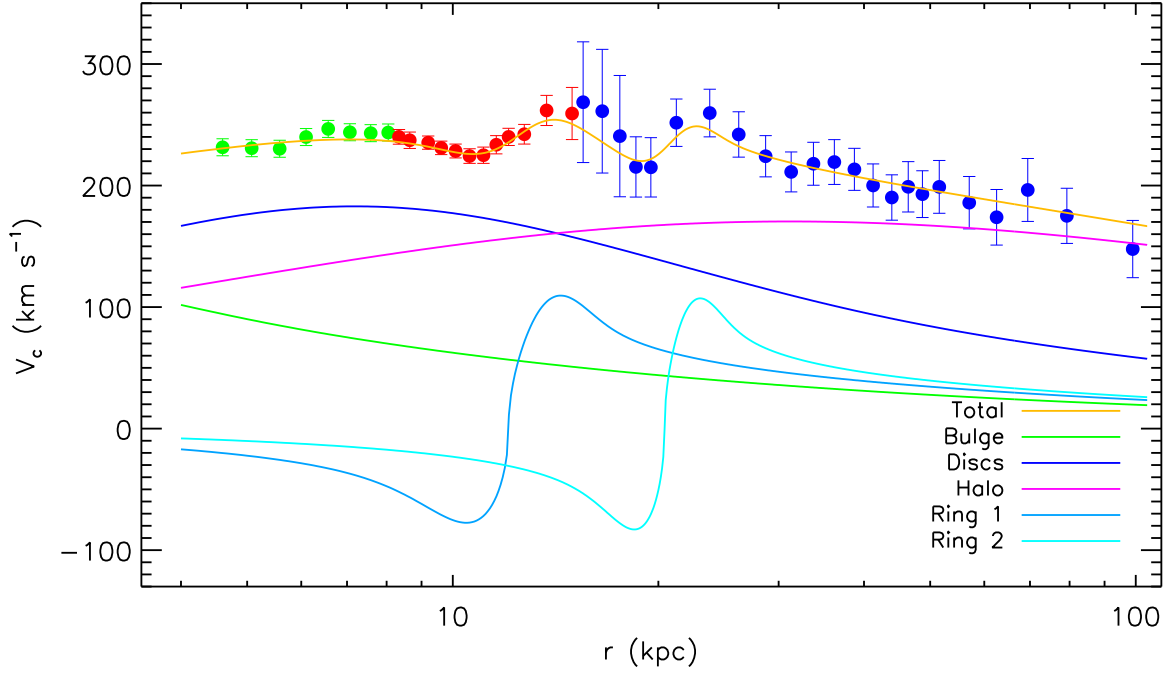}
\caption{Final combined RC of the Milky Way to $\sim$\,$100$\,kpc derived from  H~{\sc i} data (green dots), PRCGs (red dots) and HKGs (blue dots).
Lines of different colours as labeled in the bottom right corner of the diagram represent the best-fit RCs contributions to the components of the Milky Way, with the line in gold representing the sum of contributions from all the mass components (see Section\,6.1 for details).}
\end{center}
\end{figure*}

\begin{table*}
\centering
\caption{Best-fit mass model parameters and derived quantities}
\begin{threeparttable}
\begin{tabular}{lcccc}
\hline
Galactic component&Parameter&Value&Unit&Note$^{a}$\\
\hline
Bulge&$M_{\rm b}$& $8.9$&$10^{9}{\rm M}_{\odot}$ &fixed\\
\hline
discs&$\Sigma_{\rm d, 0, thin}$&$726.9_{-123.6}^{+203.5}$&${\rm M}_{\rm \odot}$\,pc$^{-2}$&fixed\\
&$R_{\rm d, thin}$&$2.63^{+0.16}_{-0.21}$& kpc&fitted\\
&$M_{\rm d, thin}$&$3.15_{-0.19}^{+0.35}$&$10^{10} {\rm M}_{\odot}$&derived\\
&$\Sigma_{\rm d, 0, thick}$&$30.4_{-10.3}^{+36.2}$&${\rm M}_{\rm \odot}$\,pc$^{-2}$&fixed\\
&$R_{\rm d, thick}$&$5.68^{+2.22}_{-1.99}$& kpc&fitted\\
&$M_{\rm d, thick}$&$0.62_{-0.06}^{+0.16}$&$10^{10}\,{\rm M}_{\odot}$&derived\\
&$\Sigma_{\rm d, 0, gas}$&$134.3_{-12.1}^{+18.8}$&${\rm M}_{\rm \odot}$\,pc$^{-2}$&fixed\\
&$R_{\rm d, gas}$&$5.26^{+0.32}_{-0.42}$ &kpc&fixed\\
&$M_{\rm d, gas}$&$0.55_{-0.02}^{+0.02}$&$10^{10}\,{\rm M}_{\odot}$&derived\\
&$M_{\rm d, total}$&$4.32^{+0.39}_{-0.20}$&$10^{10}\,{\rm M}_{\odot}$&derived\\
\hline
Dark matter halo&$r_{\rm s}$&$14.39^{+1.30}_{-1.15}$& kpc&fitted\\
&$\rho_{\rm s}$&$0.0121_{-0.0016}^{+0.0021}$ &${\rm M}_{\rm \odot}$\,pc$^{-3}$&fitted\\
&$\rho_{\rm \odot}$&$0.0083^{+0.0005}_{-0.0005}$ &${\rm M}_{\rm \odot}$\,pc$^{-3}$&derived\\
&$c$&$18.06_{-0.90}^{+1.26}$&--&derived\\
&$r_{\rm vir}$&$255.69_{-7.67}^{+7.67}$& kpc&derived\\
&$M_{\rm vir}$&$0.90_{-0.08}^{+0.07}$&$10^{12}\,{\rm M}_{\rm \odot}$&derived\\
\hline
Rings&$\Sigma_{0, \rm ring1}$&$44.89^{+13.47}_{-10.32}$&$\,{\rm M}_{\rm \odot}$\,pc$^{-2}$&fitted\\
&$R_{\rm ring1}$&$12.32^{+0.49}_{-0.37}$ &kpc&fitted\\
&$\sigma_{\rm ring1}$&$1.51^{+0.54}_{-0.45}$& kpc&fitted\\
&$M_{\rm ring1}$&$1.32^{+0.71}_{-0.50}$&$10^{10}\,{\rm M}_{\odot}$&derived\\
&$\Sigma_{0, \rm ring2}$&$27.37^{+19.16}_{-13.69}$&${\rm M}_{\rm \odot}$\,pc$^{-2}$&fitted\\
&$R_{\rm ring2}$&$20.64^{+1.03}_{-1.03}$& kpc&fitted\\
&$\sigma_{\rm ring2}$&$1.76^{+0.97}_{-0.74}$& kpc&fitted\\
&$M_{\rm ring2}$&$1.57^{+0.83}_{-0.75}$&$10^{10}\,{\rm M}_{\odot}$&derived\\
\hline
All&$M_{\rm total}$&$0.97^{+0.07}_{-0.08}$&$10^{12}\,{\rm M}_{\rm \odot}$&derived\\
\hline
\end{tabular}
\begin{tablenotes}
\item[$^{a}$]Here ``fixed'', ``fitted'' and ``derived'' denote the parameter/quantity of concern is either fixed or fitted in our mass model, or derived from the resultant model.
\end{tablenotes}
\end{threeparttable}
\end{table*}

\begin{figure*}
\begin{center}
\includegraphics[scale=0.42,angle=0]{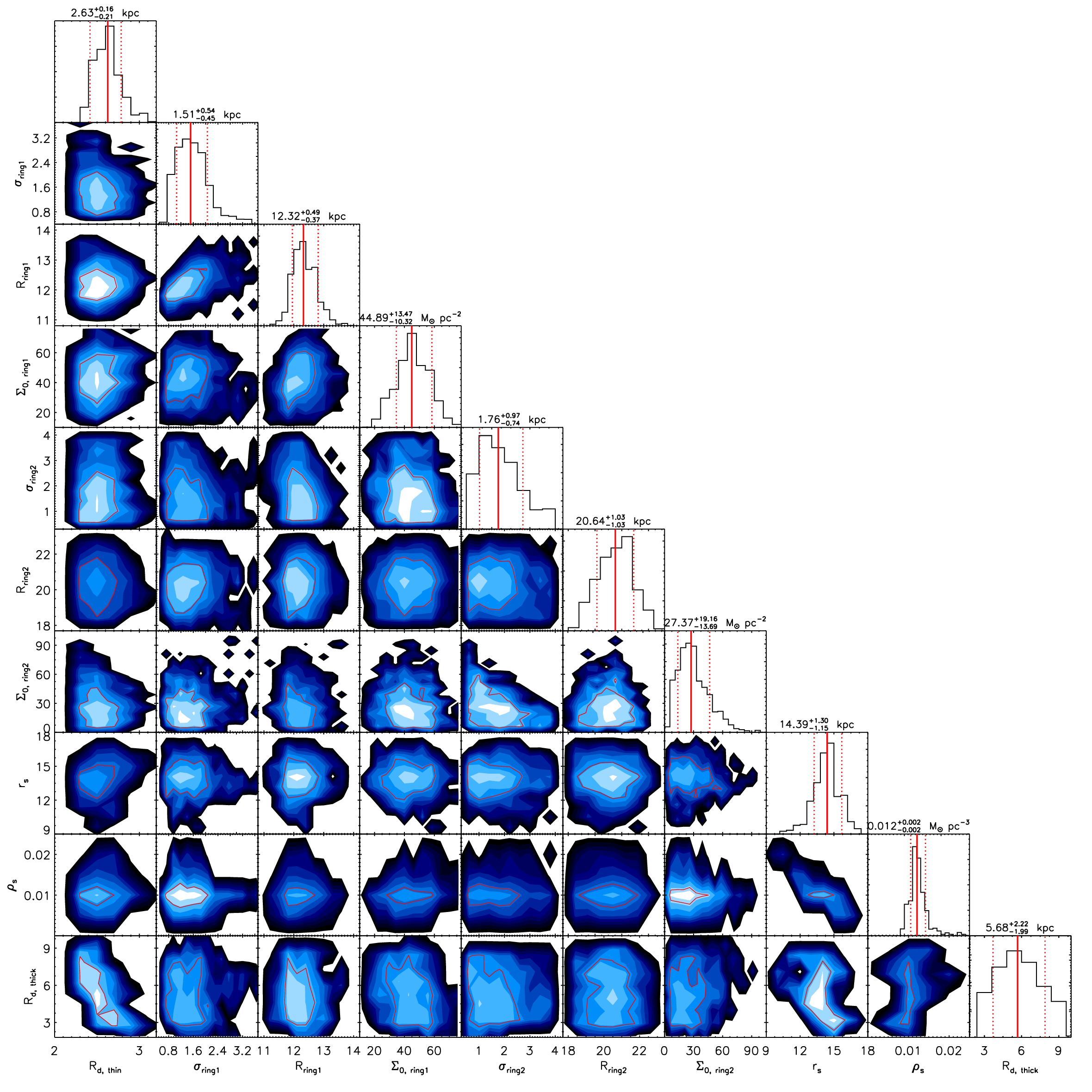}
\caption{Two-dimensional marginalized PDFs for the ten model parameters (described in detail in Section\,6.1) obtained from the MCMC.
Histograms on top of each column show the one-dimensional marginalized PDFs of each parameter labeled at the bottom of the column.
{The red contour in each panel delineates the 1-$\sigma$ confidence level.}
The red solid and dotted lines in each histogram represent, respectively,  the best-fit value and the 68 per cent probability intervals of the parameter concerned.
The best-fit values and uncertainties of the model parameters are also labeled near the top of the individual columns.}
\end{center}
\end{figure*} 

(1) \textbf{\it The discs.} Three sub-components are included  in the disc component, i.e. a gas, a thin and a thick stellar disc.
Their surface densities are all described by,
\begin{equation}
\Sigma (R) = \Sigma_{\rm d, 0}\exp(-\frac{R}{R_{\rm d}}  - \frac{R_{\rm hole}}{R} ),
\end{equation}
with a central surface density $\Sigma_{\rm d, 0}$ and a scale length $R_{\rm d}$.
The parameter $R_{\rm hole}$ is used specially for the gas disc to create a central cavity in the surface density that match with the observations (e.g. Dame 1987).
As in DB98b, we adopt $R_{\rm hole} = 4$\,kpc for the gas disc and  $R_{\rm hole} =0$ for the stellar discs. 
We further fix the surface density for the individual  sub-discs to the local measurements, denoted by $\Sigma_{R_{0}}$.
We adopt $\Sigma_{R_{0}, {\rm gas}} = 17.0\,{\rm M}_{\odot}$\,pc$^{-2}$ from Read (2014) and  $\Sigma_{R_{0}, {\rm thick}} = 7.0\,{\rm M}_{\odot}$\,pc$^{-2}$ from Flynn et al. (2006).
The local surface density of the thin disc can be derived from the total local stellar surface density  $\Sigma_{R_{0}, {\rm stellar}} = 38.0\,{\rm M}_{\odot}$\,pc$^{-2}$ estimated by Bovy et al. (2013) by subtracting contributions from the thick disc and the stellar halo ($\Sigma_{R_{0}, {\rm halo}} = 0.6\,{\rm M}_{\odot}$\,pc$^{-2}$; Flynn et al. 2006).
Then, for each sub-discs, the central surface density can be calculated from $\Sigma_{\rm d, 0} = \Sigma_{\rm R_{0}}\exp(R_{0}/R_{d} + R_{\rm hole}/R_{0})$.
Moreover, we fix the ratio of gas disc scale length to the thin disc scale length to 2.
For any axisymmetric component with surface density $\Sigma (R^{'})$, the circular velocity is given by (Binney \& Tremaine 2008; Xin \& Zheng 2013),
\begin{equation}
V_{c}^{2} (R) = -4G \int_{0}^{R} \frac{a^{2}{\rm d}a}{\sqrt{R^{2} - a^{2}}}  \int_{a}^{\infty} \frac{{\rm d}\Sigma (R^{'})}{\sqrt{{R^{'2}}-a^{2}}}.
\end{equation}
Specifically, for the exponential  { razor-thin} stellar discs, Eq.\,(13) has analytic solution given by,
\begin{equation}
 V_{c}^{2} (R) = 4\pi G\Sigma_{\rm d, 0}R_{\rm d}y^{2}[I_{0}(y)K_{0}(y) - I_{1}(y)K_{1}(y)], 
\end{equation}
where $y = R/(2R_{\rm d})$.
$I_{n}$ and $K_{n}$ ($n = 0, 1$) are the first and second kind modified Bessel functions, respectively.
Finally, the circular velocity of main disc component is given by quadratic sum of contributions from the three sub-dics,  $V_{c,\,{\rm\,disc}}^{2} = V_{c,\,{\rm\,thin}}^{2} + V_{c,\,{\rm\,thick}}^{2} + V_{c,\,{\rm\,gas}}^{2}$.

(2) \textbf{\it The bulge and dark matter halo.} The density distributions of the bulge and the dark matter halo are each described by, 
\begin{equation}
\rho (R, Z) = \frac{\rho_{0}}{m^{\gamma}(1+m)^{\beta - \gamma}} \exp[-(mr_{0}/r_{\rm t})^{2}] ,
\end{equation}
where,
\begin{equation}
m (R, Z) = \sqrt{(R/r_{0})^{2}  + (Z/qr_{0})^{2}},
\end{equation}
with scale radius $r_{0}$, scale density $\rho_{0}$, axis ratio $q$ and truncated radius $r_{\rm t}$.
The indexes $\gamma$ and $\beta$ describe respectively the inner ($r \ll r_{0}$) and outer ($r_{0} \ll r \ll r_{\rm t}$) slopes of the radial density profile.  

Following McMillan (2011), we set our bulge mass model similar to that constructed by Bissantz \& Gerhard (2002) given that we have no data to constrain the bulge mass distribution.
The model  has parameters $\gamma_{\rm b} = 0$, $\beta_{\rm b} = 1.8$, $r_{\rm b, 0} = 0.075$\,kpc, $r_{\rm b, t} = 2.1$\,kpc, an axis ratio $q = 0.5$ and a scale density $\rho_{\rm b, 0} = 9.93 \times 10^{10} {\rm M}_{\odot}$\,kpc$^{-3}$. 
The total bulge mass corresponding to these parameters is $8.9 \times 10^{9} {\rm M}_{\odot}$.
The contribution to circular velocity of this bulge component can be calculated numerically (Binney \& Tremaine 2008,  pp. 92). 
 
 For the dark matter halo, a spherical NFW density profile is adopted with $q_{\rm h} = 1$, $\gamma_{\rm h} = 1$, $\beta_{\rm h} =3$ and $r_{\rm t} \simeq \infty$.
 The free parameter $\rho_{\rm h, 0}$, also denoted as $\rho_{s}$, is given by,
 \begin{equation}
 \rho_{s} =  \frac{\rho_{\rm cr}\Omega_{\rm m}\delta_{\rm th}}{3}\frac{c^{3}}{{\rm ln}(1 + c) - c/(1 + c)},
 \end{equation}
 where  $\rho_{\rm cr} = 3H^{2}/8\pi G$  is the critical density of the universe, $\Omega_{\rm m}$ the contribution of (dark and baryonic) matter to the critical density, $\delta_{\rm th}$ the critical overdensity at virialization and $c$ the concentration parameter (the ratio of the virial radius $r_{\rm vir}$ to the scale radius $r_{h, 0}$, also denoted as $r_{s}$).
In the following analysis, we adopt $\Omega_{\rm m} = 0.28$ and $H_{0} = 69.7$\,km s$^{-1}$ Mpc$^{-1}$ from Hinshaw et al. (2013), and set $\delta_{\rm th} = 340$ (Bryan \& Norman 1998).
The enclosed virial mass within the virial radius of the NFW dark matter halo is given by, 
\begin{equation}
M_{\rm vir} =  \frac{4\pi}{3}\rho_{\rm cr}\Omega_{\rm m}\delta_{\rm th}r_{\rm vir}^{3}.
\end{equation}
The contribution to circular velocity of the NFW dark matter halo can be calculated as
\begin{equation}
V_{c}^{2} (r) = 4\pi G\rho_{s}r_{s}^{3}\frac{\ln (1 + x) - x/(1 + x)}{r},
\end{equation}
where $x = r/r_{s}$.

(3) \textbf{\it The mass rings.} As discussed earlier, there are two prominent localized dips in the RC, one at $r$\,$\sim$\,$11$\,kpc and another at $r$\,$\sim$\,$19$\,kpc.
By chance or not, the two dips are almost at the exact positions of the $n =3$ and 2  hypothetical caustic rings of dark matter, at $a_{3} \simeq 13$\,kpc and $a_{2} \simeq 20$\,kpc, respectively, as proposed by Sikivie (2003) and Duffy \& Sikivie (2008). 
To quantify the mass distributions that may be associated with the two dips,  we consider two ring-like structures in the Galactic plane in our mass models.
Following de Boer \& Weber (2011), the surface density profiles of the two rings are each described by,
\begin{equation}
\Sigma (R) = \Sigma_{\rm 0, ring}\exp[-\frac{(R - R_{\rm ring})^{2}}{2\sigma^{2}_{\rm ring}}].
\end{equation}
This Gaussian-like ring has a central surface density $\Sigma_{\rm 0, ring}$, a ring radius $R_{\rm ring}$ and a Gaussian width $\sigma_{\rm ring}$.
The contribution to circular velocity of the ring component is given by the quadratic sum of contributions of the two rings,  $V_{c, {\rm ring}}^{2} = V_{c, {\rm ring1}}^{2} +  V_{c, {\rm ring2}}^{2}$ and can be calculated numerically with Eq.\,(13) numerically.

In total, there are ten free parameters in our Galactic mass model: two for the discs ($R_{\rm d, thin}$ and $R_{\rm d, thick}$), two for the dark matter halo ($r_{s}$ and $\rho_{s}$), and six for the rings ($\Sigma_{\rm 0, ring1,2},\,$$R_{\rm ring1,2}$ and $\sigma_{\rm ring1,2}$ ).
To derive the ten free parameters, we fit the model circular velocities given by Eq.\,(11), to match our newly derived values (see Table\,3 \& Fig.\,12).
We note that the latter six parameters are only sensitive to the two localized features (i.e. the dips) in the RC and thus do not affect the overall fit controlled by the former four parameters.
To efficiently explore the parameter space in searching for the best mass model, we use a Markov Chain Monte Carlo (MCMC) technique to sample the likelihood of the data, which is defined as,
\begin{equation}
\mathcal{L} = \prod_{i=1}^{N}\frac{1}{\sqrt{2\pi}\sigma_{V_{c, R_{i}}^{\rm obs} }}\exp{\frac{-[V_{c, R_{i}}^{\rm obs} - V_{c, R_{i}}^{\rm model} (\textbf{\emph p})]^{2}}{2\sigma_{V_{c, R_{i}}^{\rm obs}}^{2}}},
\end{equation}
where $N$ is the total data points, $\sigma_{V_{c, R_{i}}^{\rm obs}}$ is the uncertainty of the observed circular velocity and \textbf{\emph p} represents the ten free parameters of the above Galactic mass model that we want to determine.
The parameters after post-burn period in the MCMC chain give the probability distribution functions (PDFs) of the ten free parameters.
We present the marginalized one- and two-dimensional PDFs of the model parameters in Fig.\,13.
The joint PDFs clearly show some correlations amongst the parameters. 
For example,   $R_{\rm ring1}$ is  strongly correlated twith $\sigma_{\rm ring1}$.
In addition, anticorrelations are found between $R_{\rm d, thin}$ and $R_{\rm d, thick}$, and between $\rho_{s}$ and $r_{s}$.
Other pairwise parameters are generally independent of each other.
Finally, the best-fit values of model parameters are estimated by the median values of their marginalized PDFs.
The uncertainties are computed from the 68 per cent probability intervals of the marginalized PDF of each parameter. 
The final best-fit values of parameters of our mass models and other derived quantities (e.g. mass of each component), together with their corresponding uncertainties, are presented in Table\,4. 
As Fig.\,12 shows, the best-fit RC is in excellent agreement with the observed one. 
The contributions to RC from each of the components corresponding to the best-fit parameters are also overplotted in Fig.\,12.

\begin{figure}
\begin{center}
\includegraphics[scale=0.42,angle=0]{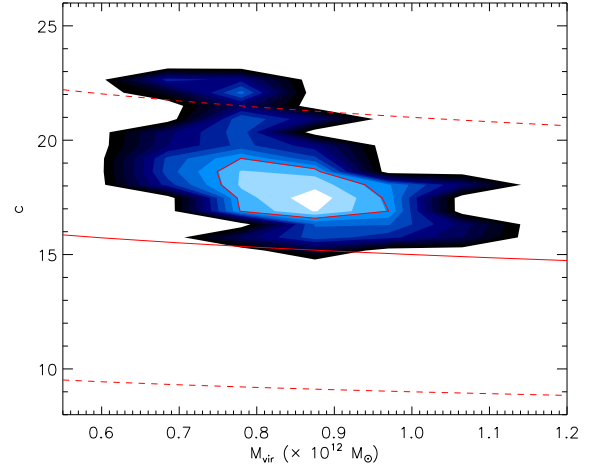}
\caption{Two-dimensional PDF of virial mass $M_{\rm vir}$ and concentration parameter $c$. The red solid line delineates the $M_{\rm vir}$--$c$ relation predicted by the $\Lambda$CDM simulations of Bullock et al. (2001).
The dashed red lines border  the scatters of the relation.
The red contour represents the 1-$\sigma$ confidence level.}
\end{center}
\end{figure}  

\subsection{Discussion}
\subsubsection{Dark matter halo}
From the best-fit values of $\rho_{s}$ and $r_{s}$, we estimate a virial mass of the dark matter halo, {$M_{\rm vir} = 0.90^{+0.07}_{-0.08} \times 10^{12} {\rm M}_{\odot}$} within the virial radius {$r_{\rm vir} = 255.69^{+7.67}_{-7.67}$\,kpc}. 
The results are in excellent agreement with the recent measurements of Kafle et al. (2014).
The concentration parameter {$c$ is $18.06^{+1.26}_{-0.90}$}, which  also agrees well with the recent determinations of Kafle et al. (2014) and Piffl et al. (2014).
As expected from Eq.\,(18), there is a strong anticorrelation between $c$ and $M_{\rm vir}$ as shown in Fig.\,14.
The $M_{\rm vir}$--$c$ joint PDF is presented in Fig.\,14. 
Also overplotted in the Figure is the relation predicted by $\Lambda$CDM simulations taken from Bullock et al. (2001).
The values of $c$ predicted by the simulations are systematically smaller than that yielded by our newly derived RC.
However, as argued by Kafle et al. (2014) and Piffl et al. (2014), the theoretical relation is constructed from  simulations with dark matter only.
The presence of baryons, not considered in the simulations, is expected to increase the concentration.
Finally, we note that the slightly lighter dark matter halo estimated here could lessen tension in hierarchical structure formation in the $\Lambda$CDM cosmological paradigm imposed by the so-called {\it missing satellite} and  {\it too big to fail} problems (e.g. Springel et al. 2008; Vera-Ciro et al. 2013). 

From $\rho_{s}$ and $r_{s}$, we also find a local dark matter density, $\rho_{\rm \odot, dm} = 0.0083^{+0.0005}_{-0.0005}$ M$_{\odot}$\,pc$^{-3}$ ($0.32^{+0.02}_{-0.02}$\,Ge\,V\,cm$^{-3}$).
The very small uncertainty (only 5 per cent) is due to the strong anticorrelation between $\rho_{s}$ and $r_{s}$.
Our estimate of $\rho_{\rm \odot, dm}$ is in good agreement with the previous global (e.g. Salucci et al.\,2010, Catena \& Ullio\,2010 and McMillan\,2011) as well as  local determinations (e.g. Bovy \& Tremaine 2012; Zhang et al. 2013; Bovy \& Rix 2013), pointing to a nearly spherical local Milky Way dark matter halo (Read 2014). 

\subsubsection{The rings}
As Fig.\,12 shows, the two localized  dips in our RC are fitted quite well by two Gaussian-like ring structures in the Galactic plane.
The best-fit values of the radii of the two rings are, respectively, {$12.32^{+0.49}_{-0.37}$\,kpc and $20.64^{+1.03}_{-1.03}$\,kpc}, in good agreement with the radii of the $n = 3$ and $2$ hypothetical caustic rings of dark matter, at  $a_{3} \simeq 13$\,kpc and $a_{2} \simeq 20$\,kpc, respectively, as proposed by Sikivie (2003) and Duffy \& Sikivie (2008).
The radius of the inner ring is also in excellent agreement with the value of $12.4$\,kpc estimated by de Boer \& Weber (2011).
The rings are quite massive, of the order of $10^{10}\,{\rm M}_{\odot}$, again matching well with estimate  of de Boer \& Weber (2011).
At present, observational evidence linking the dips seen in the RC to hypothetical caustic rings of dark matter is still marginal .
To better understand the origin of the rings (or the dips in the RC), further observations and simulations are needed. 

\section{Summary}
Based on $ 16, 000$ PRCGs selected from LSS-GAC and SDSS-III/APOGEE, and $ 5700$ HKGs selected from SDSS/SEGUE, we have derived the RC of the Milky Way out to $\sim$\,$100$\,kpc.
For the warm disk tracers PRCGs, a kinematic model with asymmetric drift correction is used to drive the RC.
Benefited from the high accuracy of distances and line-of-sight velocities of the PRCG sample, the typical uncertainties of the newly derived circular velocities are only $5$-$7$ km\,s$^{-1}$.
From the new accurate RC yielded by the PRCG sample, we have also obtained an estimate of the circular velocity of { $240\pm6$\,km\,s$^{-1}$} at the solar position, of the solar peculiar velocity in the rotation direction of {$12.1 \pm 7.6$\,km\,s$^{-1}$}.  
For the HGK halo tracers, we derive the RC by spherical Jeans equation.
We use the current available measurements of the velocity anisotropy parameter $\beta$ to break the so-called {\it RC/mass-anisotropy degeneracy}.
The typical uncertainties of the derived circular velocities are several tens km\,s$^{-1}$.

By combining circular velocities from H~{\sc i} measurements for the inner disk inside the solar circle, we present a combined RC for Galactocentric distance $r$ ranging from $\sim$\,$4$ to $\sim$\,$100$\,kpc.
The combined RC show an overall flat value of $\sim$\,$240$\,km\,s$^{-1}$ within $r$\,$\sim$\,$25$\,kpc, beyond which it declines steadily to 150\,km\,s$^{-1}$ at $r$\,$\sim$\,$100$\,kpc.
The newly derived RC also established the existence of significant localized  dips at $r$\,$\sim$\,$11$ and $\sim$\,$19$\,kpc, respectively.
The dips are possibly related to the $n = 3$ and $2$  hypothetical caustic rings of dark matter with ring radii at $a_{2} \simeq 20$\,kpc and $a_{3} \simeq 13$\,kpc, respectively, as proposed by Sikivie (2003) and Duffy \& Sikivie (2008).

Finally, we construct a parametrized Galactic mass model constrained by the newly derived  RC and other available constraints.
The best-fit yields a virial mass of the dark matter halo {$M_{\rm vir} = 0.90^{+0.07}_{-0.08} \times 10^{12}$\,${\rm M}_{\odot}$}, a concentration parameter {$c = 18.06^{+1.26}_{-0.90}$} and a local dark matter density {$\rho_{\rm \odot, dm} = 0.32^{+0.02}_{-0.02}$ GeV\,cm$^{-3}$}.
We also find the two RC dips can be well described by two Gaussian-like ring structures with the ring radii almost identical to the ones predicted by the hypothetical  $n = 2$ and $3$  caustic rings of dark matter.
The two rings have a mass of the order of $10^{10} {\rm M}_{\odot}$.

 \section*{Acknowledgements} 
 This work is supported by the National Key Basic Research Program of China 2014CB845700 and the National Natural Science Foundation of China 11473001.  
We thank Zuhui Fan and Xiangkun Liu for valuable discussions. 
We also thank Prajwal Raj Kafle for kindly providing his circular velocity measurements.  
The LAMOST FELLOWSHIP is supported by Special Funding for Advanced Users, budgeted and administrated by Center for Astronomical Mega-Science, Chinese Academy of Sciences (CAMS).

The Guoshoujing Telescope (the Large Sky Area Multi-Object Fiber Spectroscopic Telescope, LAMOST) is a National Major Scientific Project built by the Chinese Academy of Sciences. Funding for the project has been provided by the National Development and Reform Commission. LAMOST is operated and managed by the National Astronomical Observatories, Chinese Academy of Sciences.

This work has made use of data products from the Sloan Digital Sky Survey (SDSS).

\appendix
{ \section{Mock-data tests}
Here, we use mock-data to test the effects of the approximations used in our kinematical model and the performance of the whole methodology described in Section\,3 on deriving $V_{c} (R)$ and $\overline{V}_{R} (R)$.
The mock-data sets are created by re-sampling the heliocentric line-of-sight velocity, $V^{\rm helio}_{\rm los}$, for each data point from the probability distribution function (PDF)  $p(V_{\rm los}|l, b, d, R_{0}, V_{R, \odot}, V_{\phi, \odot}, V_{\phi} (R), V_{R} (R), \sigma_{R} (R))$.
The positions ($l, b, d$) of the stars in the PDF are taken as exactly as those of the real data, i.e. 15,634 PRCGs.
The values of Galactic constants ($R_{0}$, $V_{\phi, \odot}$, $V_{R, \odot}$) are the same as those  fixed in Table\,1 and the radial velocity dispersion profile, $\sigma_{R} (R)$,  is taken from the one derived in Section\,3.1 by ourselves. 
At each $R$, a Gaussian distribution is used for $V_{R}$ with a velocity dispersion, $\sigma_{R} (R)$, as just mentioned.
The centre of the Gaussian distribution, i.e.  the mean radial velocity $\overline{V}_{R}$, is a function of $R$ given by,
\begin{equation}
\overline{V}_{R} =  -10 \times \cos(2\pi R / 10 + 2).
\end{equation}
Rather than using a simplified Gaussian distribution of azimuthal-velocities assumed in our kinematical modeling described in Section\,3.1, here, we use an  azimuthal-velocity distribution generated from an analytic formula given by SB12.
As argued by SB12, the analytic formula can nicely and naturally reproduce the non-Gaussianity of the observed azimuthal velocity distribution of the Geneva-Copenhagen Survey (GCS) local sample of stars with accurate space velocities.
In addition, the distribution of $V_{\phi}$ given by this analytic formula also yields excellent fit to the distribution of $V_{\phi}$ produced by rigorous torus-based dynamics modelling (Binney \& McMillan 2011). 
The distribution of $V_{\phi}$ generated from this formula is given by,
\begin{equation}
 \begin{split}
              & n (V_{\phi} | R, z) = \mathcal{N} \exp{(-\frac{R_{\rm g} - R_{0}}{R_{d}})}\frac{2\pi R_{\rm g}K}{\sigma_{R} (R_{\rm g})}\\
              & \qquad \qquad \qquad \times \exp{[-\frac{\Delta\Phi_{\rm ad}}{\sigma_{R}^{2} (R_{\rm g})}]}f(z, R_{\rm g} - R),
 \end{split}
\end{equation}
where $\mathcal{N}$ is a normalization factor. $R_{\rm g}$ is the guiding-centre radius given by $R_{\rm g} = RV_{\phi}/V_{c}$.
$K$ is a factor that can be numerically calculated [cf. Eq.\,(12) of SB12].
$\Delta\Phi_{\rm ad}$ and $f(z, R_{\rm g} - R)$ are the the so-called adiabatic potential and $z$ factor, respectively (cf. Section\,3 of SB12 for detailed descriptions of the two terms).
The values of $R_{d}$ and $R_{0}$ are adopted from Table\,1.
The profile of radial velocity dispersion, $\sigma_{R} (R_{\rm g})$, is again the same as the one derived in Section\,3.1 by ourselves. 
For the rotation curve (RC), $V_{c} (R)$,  two types of shape are assumed and tested:
1) A flat one,
\begin{equation}
V_{c} = 220;
\end{equation}
and 2) A parabolic curve,
\begin{equation}
V_{c}  =  3 \times (R -12)^2 + 220.
\end{equation} 

Using the PDF of $V_{\rm los}^{\rm helio}$ described above, we create five mock-data sets, two with the flat and three with the parabolic RC. 
Then the same fitting technique used for the real data as described in Section\,3.2 is applied to the mock-data sets.
The RCs derived from the five mock-data sets are plotted in the left panel of Fig.\,A1.
The mean radial motions as a function of $R$ derived from the mock-data sets are also shown in the right panel of  Fig.\,A1.
For both assumed shapes of RC, the true values of circular velocity are all excellently recovered by our methodology described in Section\,3.
In addition, as we expect, the true values of the mean radial motions as a function of $R$ are also recovered quite well.
 }

\begin{figure*}
\begin{center}
\includegraphics[scale=0.65,angle=0]{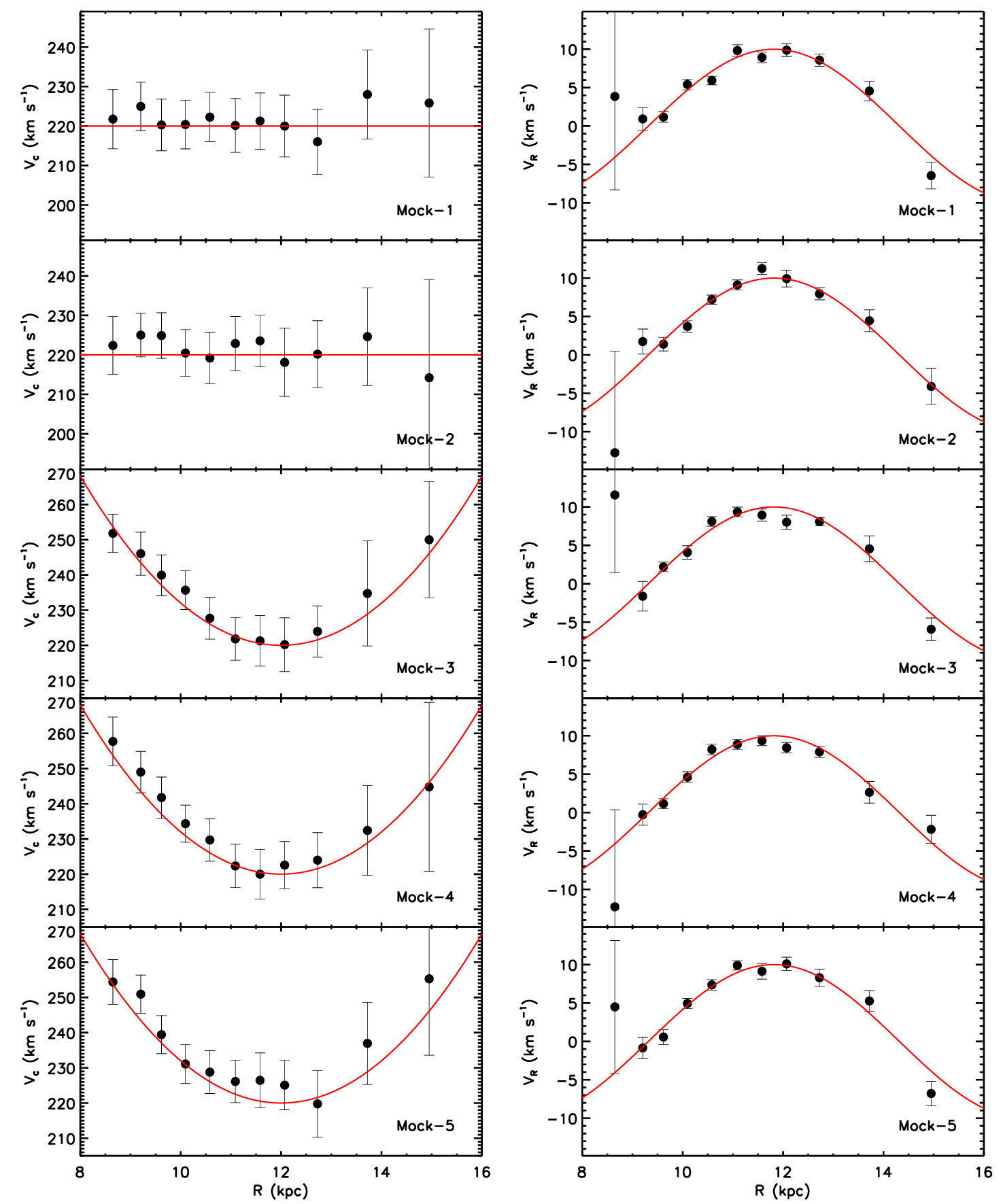}
\caption{The left and right panels show the rotation curves  and  mean radial motions, respectively, recovered from the five mock-data sets using the exactly the same fitting method as applied to the real data. 
Red lines in the left and right panels plot the true rotation curves and mean radial motions as a function of $R$, respectively, that are assumed in generating the mock-data sets.
The top two rows show the case of  mock-data sets assuming a flat rotation curve of value of 220\,km\,s$^{-1}$ [see Eq.\,(A3)], while the bottom rows are parabolic rotation curve described by Eq.\,(A4).
The true mean radial motions as a function of $R$ are described by Eq.\,(A1).
}
\end{center}
\end{figure*}  

\end{document}